\documentclass{tPHM2e_clean}

\usepackage[utf8]{inputenc}
\usepackage{subfigure}
\usepackage{setspace}
\usepackage{upgreek}
\usepackage{enumerate}
\usepackage{url}
\usepackage{color}

\DeclareMathOperator*{\argmin}{arg\,min}
\theoremstyle{plain}

\theoremstyle{definition}

\theoremstyle{remark}

\newcommand{\tightfbox}[1]{\fboxsep0pt\fboxrule0.5pt\fbox{#1}}

\begin{document}

%
%
\title{Fitting Laguerre tessellation approximations to\\ tomographic image data}

\author{
\name{
A. Spettl\textsuperscript{a}$^{\ast}$\thanks{$^\ast$Corresponding author. Email: aaron.spettl@uni-ulm.de},
T. Brereton\textsuperscript{a},
Q. Duan\textsuperscript{b},
T. Werz\textsuperscript{c},
C.\,E. Krill III\textsuperscript{c},
D.\,P. Kroese\textsuperscript{b} and
V. Schmidt\textsuperscript{a}
}
\affil{
\textsuperscript{a}Institute of Stochastics, Ulm University, 89069 Ulm, Germany;
\textsuperscript{b}School of Mathematics and Physics, The University of Queensland, Brisbane 4072, Australia;
\textsuperscript{c}Institute of Micro and Nanomaterials, Ulm University, 89069 Ulm, Germany
}
\received{Submitted 6 August 2015}
}

\maketitle

\begin{abstract}
The analysis of polycrystalline materials benefits greatly from accurate quantitative descriptions of their grain structures. Laguerre tessellations approximate such grain structures very well. However, it is a quite challenging problem to fit a Laguerre tessellation to tomographic data, as a high-dimensional optimization problem with many local minima must be solved. In this paper, we formulate a version of this optimization problem that can be solved quickly using the cross-entropy method, a robust stochastic optimization technique that can avoid becoming trapped in local minima. We demonstrate the effectiveness of our approach by applying it to both artificially generated and experimentally produced tomographic data.
\end{abstract}



\section{Introduction}\label{sec:introduction}

In recent years there have been significant advances in the tomographic characterization of materials. As a result, it is now possible to carry out detailed investigations of the 3D grain structures of polycrystalline materials; see, e.g., \cite{Doebrich04,Rowenhorst06,Limodin07,Groeber08a,Ludwig09,Lyckegaard10,Werz14}. A fundamental ingredient in any such investigation is a suitable quantitative description of the grain morphology. Such a description contains the key features of the structure, ideally free from noise and imaging artifacts. A good description usually results in significant data compression, describing large 3D voxel data sets using only a small number of parameters. Data compression is necessary, for example, when carrying out analysis of sequences of tomographic data sets (e.g., the high time resolution in-situ synchrotron images considered in \cite{Spettl15b}). In addition, the description of tomographic data via tessellations provides a basis for the statistical analysis of grain structures and, in some cases, can be used to develop stochastic models of material microstructures; see, e.g., \cite{Fan04,Beil06,Groeber08b,Lautensack08a,Lautensack08b,Schmidt15,Redenbach15,Spettl15b, Westhoff15}.

The most commonly used quantitative descriptions of space-filling grain ensembles are based on tessellations, which divide the space into disjoint regions called {\em cells}. The cells represent the individual grains. The most widely used tessellation model is the {\em Voronoi tessellation} (see, e.g., \cite{Chiu13, Okabe00}), which takes, as parameters, a set of generating points. The space is then divided into convex cells by assigning each point to its nearest generator. The {\em Laguerre tessellation} (see, e.g., \cite{Chiu13, Okabe00}) is a generalization of the Voronoi tessellation that also partitions the space into convex cells. Like the Voronoi tessellation, the Laguerre tessellation is generated by a set of points; however, unlike the Voronoi tessellation, these points are weighted, with the weights influencing the size of the cells. Consequently, the Laguerre tessellation is able to describe a wider range of structures than the Voronoi tessellation. For this reason, the Laguerre tessellation is a popular choice for modeling polycrystalline grain structures \cite{Fan04,Telley92,Telley96a,Telley96b,Xue97,Spettl15b} and other materials, such as foams \cite{Lautensack06,Lautensack08a,Liebscher14}.  

In order to describe a tessellation by a set of generating points, it is necessary to solve an inverse problem: that is, a set of generating points that produce the observed cells must be found. The {\em Voronoi inverse problem} (VIP) is well-studied; see, for example, \cite{Ash85,Evans87,Aurenhammer87b,Hartvigsen92,Schoenberg03,Adamatzky93}. Recently, Duan et al. \cite{Duan14} proposed an algorithm that finds solutions to the {\em Laguerre inverse problem} (LIP). Although the examples considered in \cite{Duan14} are restricted to 2D, the methodology is easily applied in higher dimensions.

The solutions to the VIP and the LIP assume that the empirical data constitute perfect descriptions of the observed cells. However, this is not true when working with tomographic data, which is distorted by noise and also contains imprecision arising from discretization during the imaging process. It is also worth noting that real-world materials are not perfectly described by Laguerre tessellations (though the descriptions can be quite good). These limitations mean that methods that attempt to invert a tessellation extracted from the tomographic data do not, in general, result in good fits. The LIP is solved by iteratively finding the generating points of the given tessellations. When applied to imperfect data, this iterative procedure propagates errors, resulting in tessellations that have little correspondence to the tomographic data. Thus, when dealing with empirical data, it is not appropriate to attempt to solve the LIP. Instead, the generating points of a Laguerre tessellation that is a good approximation of the material must be found. This is, at its core, an optimization problem. We call this problem the {\em Laguerre approximation problem} (LAP). The corresponding Voronoi approximation problem has been considered in the literature, beginning with \cite{Suzuki86}.

A simple heuristic approach for solving the LAP was proposed in \cite{Lyckegaard10}. More sophisticated approaches, which formulate and solve an optimization problem, are described in \cite{Liebscher15,Alpers15,Liebscher15b}. Although these techniques provide good fits in certain settings, they are either limited to small sample sizes or require the considered tessellations to be sufficiently regular.

In this paper, we present a fast and robust algorithm for fitting Laguerre approximations to large data sets. More precisely, we formulate an optimization problem where we minimize the discrepancy between the grain boundaries observed in the image data and the grain boundaries produced by our Laguerre approximation. The cost function is chosen so that it can be evaluated very efficiently and that all necessary information can be easily obtained from image data. We then solve the optimization problem using the cross-entropy (CE) method \cite{Rubinstein04,Kroese06,Kroese11,Botev13}, a stochastic optimization algorithm that is able to escape local minima. We carry out experiments on both real and artificially-generated image data that show our approach is able to produce very good fits. 

This paper is structured as follows. In Section~\ref{sec:laguerre}, we review some key properties of Laguerre tessellations. In Section~\ref{sec:optimization}, we give a more complete description of the LAP and formulate our optimization problem. Then, in Section~\ref{sec:ce-method}, we introduce the CE method as a robust tool for solving this optimization problem. Section~\ref{sec:results} gives results for both artificial and experimental data that demonstrate the effectiveness of our approach. Finally, Section~\ref{sec:conclusions} summarizes our results and suggests directions for further research.

\section{The Laguerre tessellation}\label{sec:laguerre}

In the following section, we define Voronoi and Laguerre tessellations and give a number of properties that we will use to solve the LAP. For notational convenience, we only consider tessellations in $\mathbb{R}^3$. However, our methods are easily applied in other dimensions. 

The Voronoi tessellation is defined by a locally finite set of generating points, $\{\mathbf{x}_i\}_{i \in \mathcal{I}}$, where $\mathcal{I} \subseteq{\mathbb{N}}$ denotes the index set consisting of natural numbers. The cell corresponding to the $i$th generating point, $\mathbf{x}_i \in \mathbb{R}^3$, is given by
\begin{equation}
\label{eq:vor_def}
C_i = \{ \mathbf{y} \in \mathbb{R}^3 : \| \mathbf{y} - \mathbf{x}_i \| \leq  \| \mathbf{y} - \mathbf{x}_j \|  \text{ for all } j \in \mathcal{I}\}\,,
\end{equation}
where $\| \cdot \|$ is the Euclidean norm on $\mathbb{R}^3$. In the case of the Laguerre tessellation, the generating points, $\{(\mathbf{x}_i, r_i)\}_{i \in \mathcal{I}}$, are weighted, where we assume that $r_i>0$.
The cells of the Laguerre tessellation, $\{C_i\}_{i \in \mathcal{I}}$, are then defined by
$$
C_i = \{ \mathbf{y} \in \mathbb{R}^3 : \operatorname{pow}(\mathbf{y}, (\mathbf{x}_i,r_i)) \leq  \operatorname{pow}(\mathbf{y}, (\mathbf{x}_j,r_j))  \text{ for all } j \in \mathcal{I}\} \,,
$$
for all $i \in \mathcal{I}$, where  the Euclidean norm used in \eqref{eq:vor_def} is replaced by the so-called power distance
$$
\operatorname{pow}(\mathbf{y}, (\mathbf{x},r)) = \|\mathbf{y}-\mathbf{x}\|^2 - r^2\,.
$$
Given the generator points, the faces of the cells can be computed efficiently; see, e.g., \cite{Aurenhammer87a,Sugihara00,Imai85}. 
As we take the weights $r_i$ to be positive real numbers, the generating points have a geometric interpretation as spheres. That is, we can represent a generating point, $(\mathbf{x}_i, r_i)$, as a sphere with center $\mathbf{x}_i$ and radius $r_i$.

The flexibility of the Laguerre tessellation comes at a cost. For example, while each cell of a Voronoi tessellation contains its generating point, the generating points of a Laguerre tessellation may not be contained in their corresponding cells. In some cases, a generating point of a Laguerre tessellation may not even produce a cell. That is, it is possible that there exists a point, $(\mathbf{x}_i, r_i)$, such that $C_i = \emptyset$; see, for example, \cite{Aurenhammer87a}. In addition, while the cells of a Voronoi tessellation uniquely determine its generating points, there are uncountably many sets of generating points that can generate a given Laguerre tessellation; see, \cite{Lautensack07,Aurenhammer87c,Duan14}. Thus, while the VIP has a unique solution, the LIP has uncountably many solutions. Under mild conditions, however, it can be shown that the generating points of a Laguerre tessellation are uniquely determined given one generating point and the weight (or one coordinate) of a generating point in an adjacent cell; see \cite{Duan14}. 


We will often find it convenient to consider planes that are equidistant from two generating points (under their respective power distances). That is, given two generating points, $(\mathbf{x}_i, r_i)$ and $(\mathbf{x}_j, r_j)$, we consider the separating plane given by
$$
P_{i,j} = \{ \mathbf{y} \in \mathbb{R}^3: \mathbf{n}_{i,j}^\top \mathbf{y} + b_{i,j} = 0\} \,,
$$
where 
$$
\mathbf{n}_{i,j} = \frac{2 \cdot (\mathbf{x}_j-\mathbf{x}_i)}{\| 2 \cdot (\mathbf{x}_j-\mathbf{x}_i) \|}
$$
is the unit normal vector of $P_{i,j}$ and 
$$
b_{i,j} = \frac{\| \mathbf{x}_i \|^2 - \|\mathbf{x}_j\|^2 + r_j^2 - r_i^2}{\|2 \cdot (\mathbf{x}_j-\mathbf{x}_i)\|}
$$ 
is the shortest distance from the origin to the plane; see, e.g., \cite[Section 2.1]{Lautensack07} for more details. The plane $P_{i,j}$ defines a half-space 
$$
H_{i,j} = \{ \mathbf{y}\in\mathbb{R}^3 : \mathbf{n}_{i,j}^\top \mathbf{y} + b_{i,j} \leq 0\} \, ,
$$
which covers the cell $C_i$. Note that the intersection of all such half-spaces, $\bigcap_{k\in \mathcal{I} \setminus\{i\}} H_{i,k}$, defines the cell $C_i$. Because the normal vectors are taken to be unit vectors, the distance from an arbitrary point, $\mathbf{x} \in \mathbb{R}^3$, to the plane $P_{i,j}$ is given by
\begin{equation}
\label{eq:plane_dist}
 d(\mathbf{x}, P_{i,j}) = \vert \mathbf{n}^\top_{i,j} \mathbf{x} + b_{i,j} \vert \,. 
\end{equation}

\section{The Laguerre approximation problem (LAP)}\label{sec:optimization}

Given an exact description of a Laguerre tessellation (e.g., in terms of the half-spaces defined above), it is not too difficult to solve the LIP. That is, it is straightforward to find a set of weighted points that is able to generate the given tessellation. Furthermore, these points can be chosen to satisfy certain constraints; see \cite{Duan14}. When dealing with tomographic data, however, the description of the tessellation is not exact. This is because, even if the material itself can be perfectly described by a Laguerre tessellation, the noise and discretization errors inherent in the imaging process mean that the cell boundaries extracted from the data are subject to error.
The LIP can be solved as described in \cite{Duan14}. However, the iterative computation of generators is sensitive to imperfect data and errors propagate quickly. As a result, the ensuing tessellations often do not correspond well to the tomographic data. Therefore, solving the LIP for empirical data is generally ill advised.
Instead, we wish to find generating points that produce a Laguerre tessellation that is as close as possible to the tomographic data (with respect to some metric). This is, at its core, an optimization problem. In order to properly formulate this optimization problem, a discrepancy measure needs to be defined. We then choose the generating points of the approximating tessellation in order to minimize this discrepancy. The choice of discrepancy measure depends on the nature of empirical data. In this paper, we work with tomographic data.

\subsection{Tomographic image data}
\label{sec:data}

\begin{figure}[htb]
\begin{center}
\includegraphics[width=0.45\textwidth]{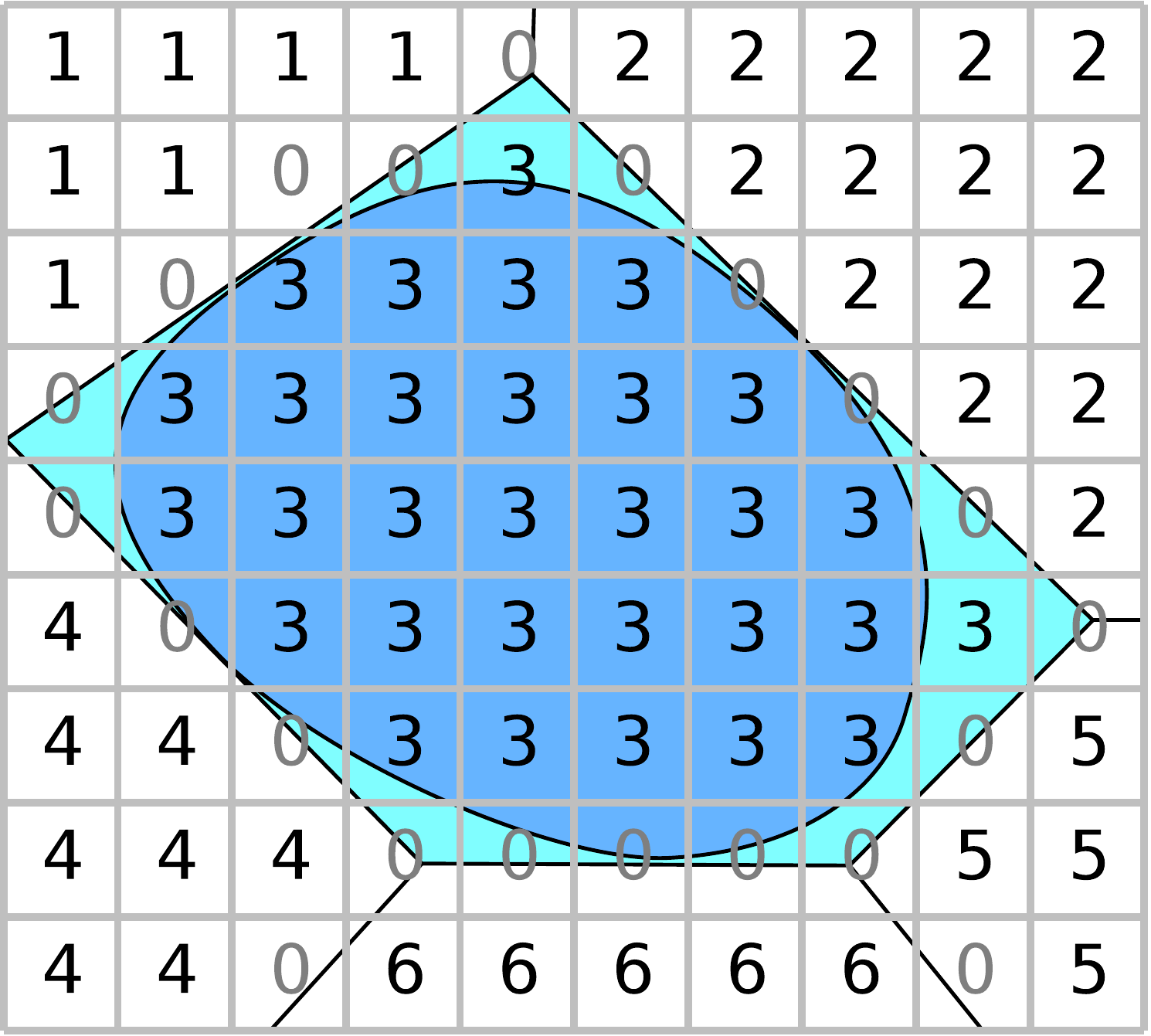}
\caption{Illustration of voxels labeled by their corresponding grains. The grain regions are assumed to form a tessellation with convex cells. The grains themselves may not fill their corresponding regions completely, as depicted for grain number 3. Here, we show the grain region (i.e., voxels with label 3), a cell approximating the grain region (colored cyan) and the grain itself (colored blue).}
\label{fig:labeled-image}
\end{center}
\end{figure}

We assume that the tomographic image data constitutes a collection of voxels in a convex window. Furthermore, we assume that the image has already been segmented --- e.g., by the watershed transform; see \cite{Roerdink00,Brunke05} --- and that the voxels have been labeled by their corresponding grains. Thus, the data is of the form $\{ I(x,y,z) \in \{0,\ldots,N\} : (x,y,z) \in W \}$, where $N\geq 1$ denotes the number of grains and $W \subset \mathbb{N}^3$ is a grid of voxel coordinates. The $i$th grain region, $R_i$, is then given by $R_i = \{ (x,y,z) \in W : I(x,y,z) = i \}$. Note that $R_0$ does not correspond to an actual grain, but is either the empty set or a collection of one voxel thick layers that separate grains. Such thin layers often arise in segmentation procedures such as the watershed transformation. A grain region $R_i$ may correspond to the grain itself or a region that contains the grain (if the space is not completely filled with grains), cf. Figure~\ref{fig:labeled-image}. We assume that the grain regions are roughly convex and that the segmentation is of a high quality.

\subsection{Discrepancy measures}

The most direct way to measure the discrepancy between 3D image data and a Laguerre tessellation generated by the points $\{(\mathbf{x}_i, r_i)\}_{i \in \mathcal{I}}$ is by counting the number of incorrectly assigned voxels. That is, we calculate
$$
\mathcal{D}_{\mathsf{vox}}\left(\{(\mathbf{x}_i,r_i)\}_{i \in \mathcal{I}} \right) = \# \left\{(x,y,z) \in W : I(x,y,z) > 0, I(x,y,z) \neq \widetilde{I}(x,y,z) \right\},
$$
where $\widetilde{I}$ is a discretized version of the tessellation generated by $\{(\mathbf{x}_i,r_i)\}_{i \in \mathcal{I}}$ and $\# A$ denotes the number of elements in some set $A$. Many other discrepancy measures considered in the literature are of a similar form. For example, in \cite{Liebscher15b}, the difference between a cell in the approximating tessellation and its equivalent in the empirical data is measured by the intersection of the approximating cell with the corresponding adjacent cells in the empirical data. Minimizing the total overlap of the approximating cells is equivalent to minimizing the number of incorrectly assigned voxels. We call discrepancy measures that aim to minimize the volume of the difference between the empirical and approximating tessellations {\em volume-based} measures.

Although minimizing a volume-based measure gives a very good fit to the data, such measures are poorly suited to most numerical optimization methods. This is because evaluating such a discrepancy measure is computationally expensive. For example, when considering the number of incorrectly assigned voxels, a discretized version of the approximating tessellation needs to be generated. In practice, this limits the number of times such a discrepancy measure can be evaluated. As a result, approaches that aim to minimize volume-based discrepancies are restricted to small data sets with small numbers of grains (e.g., 109 grains in \cite{Alpers15}) or are forced to use non-optimal optimization techniques, such as gradient-descent, which do not require too many evaluations of the discrepancy and, as such, limit the time taken by the fitting procedure (ideally to substantially less than 24 hours).

As we propose to use stochastic optimization techniques to solve this problem, we need a discrepancy measure that is significantly faster to evaluate. We can achieve this by considering an {\em interface-based} discrepancy measure --- a measure that considers only the boundaries between cells --- instead of a volume-based one. If an approximating tessellation can accurately reproduce these interfaces, it will be a good approximation of the data. The primary advantage of interface-based discrepancy measures is that they can be calculated from the generating points of the approximating tessellation without the need to generate the tessellation itself. 

In order to define such a discrepancy measure, we consider the sets of voxels that separate adjacent cells. We define the interface between two grains, $i$ and $j$, in the empirical data by 
$$
N_{i,j} = \{ (x,y,z) \in W : \mathcal{N}_{26}(x,y,z) \cap R_i \neq \emptyset \text{ and } \mathcal{N}_{26}(x,y,z) \cap R_j \neq \emptyset \} \,, 
$$
where $\mathcal{N}_{26}(x,y,z)$ denotes the 26-neighborhood of $(x,y,z)$ --- i.e., the voxels $(x^\prime,y^\prime,z^\prime)$ that have a distance less than or equal to $\sqrt{3}$ from $(x,y,z)$. This set contains all voxels that touch both grains. Note that $N_{i,j} = \emptyset$ if the grains are not adjacent. If the tessellation generated by $\{(\mathbf{x}_i, r_i)\}_{i \in \mathcal{I}}$ is a good approximation to the empirical data, then the plane separating the generating points of adjacent cells $i$ and $j$ (which defines the boundary of the two cells) should be close to $N_{i,j}$. Thus, we measure the distance between $N_{i,j}$ and $P_{i,j}$, the plane separating cells $i$ and $j$ in the approximating tessellation. The discrepancy between the approximating tessellation and the empirical data is then given by the sum of squares of these distances. That is, we define 
\begin{equation}
\label{eq:cal_D}
\mathcal{D} \left( \{ (\mathbf{x}_i,r_i) \}_{i \in \mathcal{I}} \right) = \sum_{j=1}^N \, \sum_{k=j+1}^N \, \sum_{\mathbf{y} \in N_{j,k}} d(\mathbf{y}, P_{j,k})^2,
\end{equation}
where $d(\mathbf{y}, P_{j,k})$ is given in \eqref{eq:plane_dist} and $\{P_{j,k} \}_{j,k \in \mathcal{I}}$ is the set of separating planes determined by the generating points $\{ (\mathbf{x}_i,r_i) \}_{i \in \mathcal{I}}$. Note that this discrepancy measure can be calculated without generating the approximating tessellation. A separating plane can be computed for all pairs of generators (or generator candidates, when performing the optimization) --- even if their cells are empty or not adjacent in the Laguerre tessellation. Therefore, such `degenerate' cases are not a problem for our approach. By matching all separating planes to the corresponding test points at once, Laguerre cells computed based on `good' configurations of generators will match the desired cell configuration well enough.

\subsection{Minimizing the discrepancy}

When aiming to minimize a volume-based discrepancy, the LAP reduces to the optimization problem of finding 
\begin{equation}
\label{eq:opt_prob}
\{(\mathbf{x}^*_i,r^*_i) \}_{i \in \mathcal{I}} = \argmin_{\{ (\mathbf{x}_i,r_i) \}_{i \in \mathcal{I}}} \mathcal{D}_{\mathsf{vox}} \left( \{ (\mathbf{x}_i,r_i) \}_{i \in \mathcal{I}}\right).
\end{equation}
There are a number of significant difficulties which must be overcome in solving \eqref{eq:opt_prob}. In particular,
\begin{enumerate}[(i)]
\item the optimization problem is high-dimensional,
\item the optimization problem has many local minima,
\item the discrepancy is expensive to evaluate.
\end{enumerate}

The approach developed in \cite{Alpers15} avoids problems (ii) and (iii) by finding the exact solution of a linear program (an optimization problem where a linear cost function is minimized subject to linear constraints). However, when the LAP is transformed into a linear program, the size of the resulting problem grows very quickly in both the number of voxels and the number of grains. This means that the linear programming approach cannot be applied to large 3D data sets containing many grains. 

In \cite{Liebscher15b}, gradient-descent methods are used to obtain Laguerre approximations (with a relatively low number of evaluations of the discrepancy). Unlike the linear programming approach, the dimensionality of the optimization problem grows only linearly in the number of grains (and does not depend on the number of voxels). Thus, \cite{Liebscher15b} avoids problem (iii) and, to a lesser extent, problem (i). However, the gradient-descent methods used there become stuck in local minima. Thus, the approach considered in \cite{Liebscher15b} is reliant on the ability to find initial conditions that are close to global optima. Although it is often possible to find good initial conditions --- e.g., when fitting tessellations to foams with regular structures --- this is not always the case, as will be seen in Section \ref{sec:results}. In addition, even when good initial conditions can be found, the quality of the approximating tessellation can usually be significantly improved when the optimization algorithm is able to escape local minima. 

Stochastic optimization methods are widely used tools for solving high-dimensional optimization problems with many local minima; see \cite{Kroese11} for an overview. Such methods have been used to solve problems related to the LAP. In \cite{Duan14}, the CE method is used to find a solution to the LIP and, in \cite{Liebscher15}, simulated annealing is used to fit 3D Laguerre tessellations to 2D image data. However, no stochastic approach has yet been proposed in order to solve the LAP for large, noisy 3D data sets. 

In our approach to solving the LAP for large tomographic data sets, we use the CE method to minimize an interface-based discrepancy measure. The CE method has been widely applied to high-dimensional multi-extremal optimization problems; see, \cite{Rubinstein04,Kroese06,Kroese11,Botev13}. It has a number of advantages over other stochastic optimization methods. For example, simulated annealing (described in \cite{Kirkpatrick83,Kroese11}) cannot be easily applied to the LAP, as it is very difficult to find an appropriate cooling schedule. In addition, the CE method can be easily parallelized.

By using the CE method to minimize an interface-based discrepancy measure, we have a method that can escape local minima and solves a problem whose dimensionality grows linearly in the number of grains. Although the discrepancy is expensive to evaluate, we reduce this cost significantly by minimizing an interface-based measure. In addition, we further reduce the cost of calculating the discrepancy by replacing $\mathcal{D}$ defined in \eqref{eq:cal_D} with an approximation, $\widetilde{\mathcal{D}}$.

\subsection{Approximating the discrepancy}
\label{sec:approx}

The interface-based discrepancy, $\mathcal{D}$, measures the distance between voxels on the boundaries between cells in the empirical data and the planes separating the generating points in the approximating tessellation. In order to calculate this discrepancy, every voxel in each grain boundary is considered. Although this is much faster to calculate than a volume-based discrepancy, it is still computationally expensive. However, very good approximations of $\mathcal{D}$ can be obtained by instead considering sets of {\em test points} that describe the interfaces between the grains. We find that test points obtained by fitting approximating planes to the interfaces, using the orthogonal regression approach introduced in \cite{Spettl14}, work very well.
  
In order to calculate the test points, we consider only the points separating the two cells being considered (i.e., we ignore points of contact between three or more grains). Thus, instead of $N_{i,j}$, we consider
$$
N_{i,j}^\star = \{ (x,y,z) \in N_{i,j} : (x,y,z) \notin N_{i,k} \text{ and } (x,y,z) \notin N_{k,j}, k \in \{1,\ldots,N\}, k\notin\{i,j\} \}  \,.
$$
This avoids some numerical issues that could arise when using orthogonal regression.

\begin{figure}[htb]
\begin{center}
\includegraphics[width=0.6\textwidth]{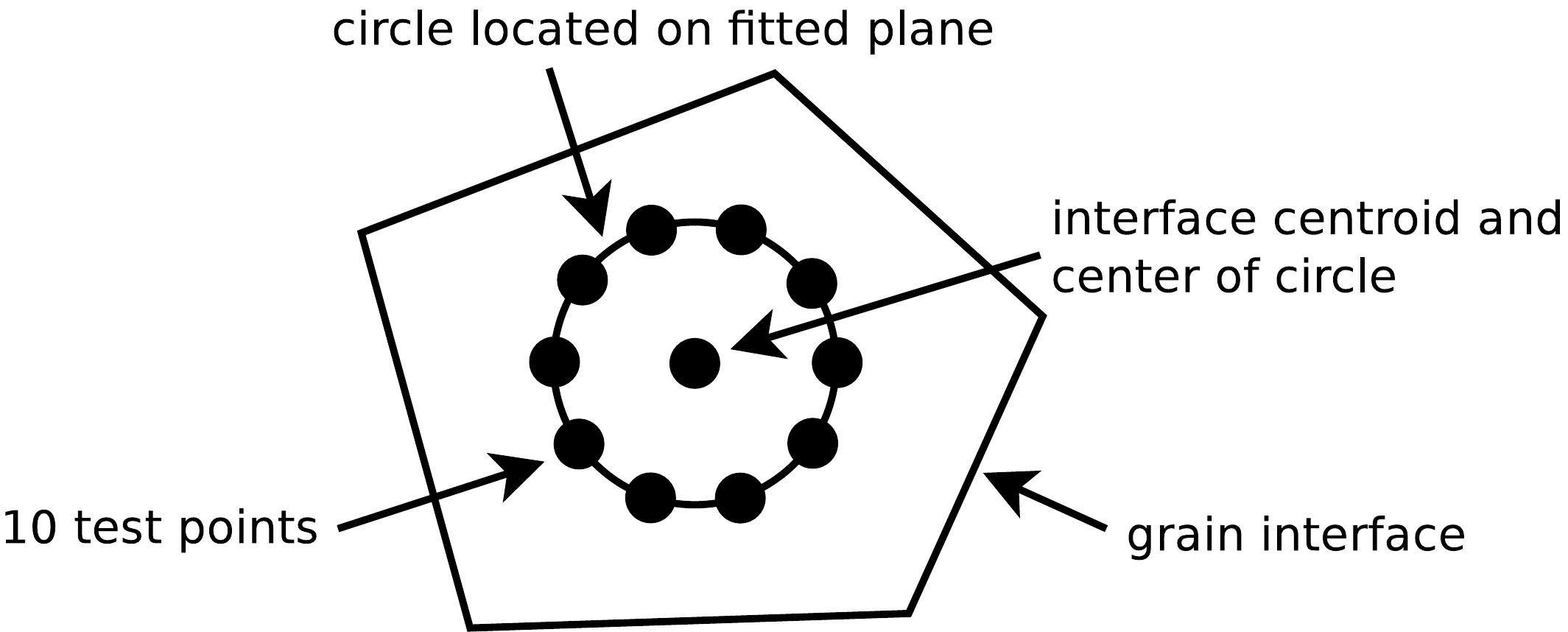}
\caption{Illustration of test points construction. First, a circle is fitted to a grain interface. The center of the circle is given by the centroid of the voxels adjacent to both grains, its 3D orientation is obtained from the orthogonal regression plane. The radius is chosen with respect to the interface area. Finally, 10 test points are placed equidistantly on this circle.}
\label{fig:test_points_on_plane}
\end{center}
\end{figure}

We then approximate the boundary between the $i$th and $j$th grain using the plane that minimizes the total squared distance to all points in $N^\star_{i,j}$. Determining this plane is a least-squares problem, which we solve using singular value decomposition. The approximating plane passes through the centroid, $\mathbf{c}\in\mathbb{R}^3$, of the set $N_{i,j}^\star$. We obtain a normal vector, $\mathbf{n}\in\mathbb{R}^3$, to the plane by taking the right-singular vector corresponding to the smallest singular value of the matrix containing the voxel coordinates in $N_{i,j}^\star$ shifted by $-\mathbf{c}$. For more details on the singular value decomposition approach, see \cite{Groen96}. The test points, $T_{i,j} \subset \mathbb{R}^3$ are chosen to lie on this approximating plane. More precisely, we put a circle with radius $\sqrt{\# N_{i,j}^\star}/4$ around the centroid $\mathbf{c}$ with the same orientation as the plane. We then place $10$ test points equidistantly on this circle. An illustration is given in Figure~\ref{fig:test_points_on_plane}.

In cases where the estimation of the plane is not `stable' (i.e., the number of voxels in $N_{i,j}^\star$ is too low to determine the location and orientation), we simply use the centroid $\mathbf{c}$ for all 10 test points. The criterion for `stable' is simple: the smallest singular value must be smaller than half the second largest one. This way we can be confident that the third and shortest axis (which is perpendicular to the plane) is correctly identified.

Given the test points for the boundary between adjacent grains $i$ and $j$, we define the approximate discrepancy at the boundary by
$$
\widetilde{\mathcal{D}}_{i,j} \left(\mathbf{x}_i, r_i, \mathbf{x}_j, r_j \right) = \sum_{\mathbf{y} \in T_{i,j}} d(\mathbf{y}, P_{i,j})^2 \,,
$$
where $d(\mathbf{y}, P_{j,k})$ is given in \eqref{eq:plane_dist} and $P_{i,j}$ is the plane separating the generating points $(\mathbf{x}_i, r_i)$ and $(\mathbf{x}_j, r_j)$. The total approximate discrepancy is then given by 
\begin{equation}
\label{eq:disc_approx}
\widetilde{\mathcal{D}} \left( \{ (\mathbf{x}_i,r_i) \}_{i \in \mathcal{I}} \right) = \frac{1}{\sum_{j=1}^N \, \sum_{k=j+1}^N \# T_{j,k}} \sum_{j=1}^N \, \sum_{k=j+1}^N \widetilde{\mathcal{D}}_{j,k} \left(\mathbf{x}_j, r_j, \mathbf{x}_k, r_k \right).
\end{equation}
Note that, here, we normalize the discrepancy. This is done in order to make the cost function easier to interpret: it can be thought of as the average squared distance of test points from their corresponding separating plane.

\section{Solving the LAP using the CE method}\label{sec:ce-method}

\subsection{The CE method}
\label{sec:CE_description}
The cross-entropy (CE) method is a stochastic optimization method that is able to solve many difficult optimization problems, including combinatorial optimization problems and continuous optimization problems with many local minima; see \cite{Rubinstein04, Botev13}. The fundamental idea of the method is to describe the location of the global minimum of an $m$-dimensional cost function, $S(\cdot)$, in terms of a degenerate $m$-dimensional probability distribution. That is, a random variable with this distribution takes only a single value, namely $\mathbf{x}^* = \argmin_{\mathbf{x}\in\mathbb{R}^m} S(\mathbf{x})$. If there is more than one global minimum, each minimum will have a corresponding degenerate distribution. The CE method works by `learning' one of these distributions. In the continuous setting, this is done as follows. A parametric density, $f(\mathbf{x};\boldsymbol{\theta})$, is used to describe the possible locations of a global minimum. A sample of size $M$ is drawn from this density and ordered by the corresponding values of $S$. The $\lceil \rho M \rceil$ (where $\lceil \cdot \rceil$ is the ceiling function) members of the sample with the lowest values of $S$ are identified as the `elite' sample, where $\rho \in (0,1)$. The elite sample is then used to update the parameter vector, $\boldsymbol{\theta}$. The updating is done by choosing $\boldsymbol{\theta}$ to minimize the cross-entropy distance (also called the Kullback-Leibler divergence) between $f(\mathbf{x};\boldsymbol{\theta})$ and the targeted degenerate probability distribution. The process is continued until a stopping condition is met, e.g., the probability distribution is nearly degenerate. Thus, the CE method uses information about good choices of arguments to find better choices. The CE method and its convergence properties are discussed extensively in \cite{Rubinstein04,Kroese06,Kroese11,Botev13}.

Usually, the parametric density, $f(\mathbf{x}; \boldsymbol{\theta})$, is chosen to be a product of normal densities. That is, we choose 
$$
f(\mathbf{x}; \boldsymbol{\theta}) = \varphi(x_1; \mu_1, \sigma_1) \varphi(x_2; \mu_2, \sigma_2) \cdots \varphi(x_m; \mu_m, \sigma_m),
$$
where $\varphi(\cdot; \mu, \sigma)$ is the density of a normal distribution with mean $\mu$ and standard deviation $\sigma$ and $\boldsymbol{\theta} = (\mu_1, \sigma_1, \ldots, \mu_m, \sigma_m)$. Thus, each component of the argument of $S(\cdot)$ is drawn independently from a normal distribution. There are two main reasons for using normal distributions. First, as the standard deviation of a normal distribution goes to zero, the distribution converges to a degenerate distribution. The second reason is that the parameter vector, $\boldsymbol{\theta}^*$, which minimizes the cross-entropy distance, is simply given by the maximum-likelihood estimates obtained from the elite sample. In other words, we update the parameters by setting the means equal to the sample means of the elite sample and the standard deviations equal to the sample standard deviations of the elite sample.

Using this approach, we have the following general algorithm for estimating $\mathbf{x}^* = \argmin_{\mathbf{x} \in \mathbb{R}^m} S(\mathbf{x})$.
\begin{enumerate}[1)]
\item \textbf{Initialization.} Identify an initial parameter vector, \newline $\boldsymbol{\theta}^{(0)} = (\mu^{(0)}_1, \sigma^{(0)}_1, \ldots, \mu^{(0)}_m, \sigma^{(0)}_m)$ and set $n = 0$.
\item \textbf{Sampling.} Draw an independent sample $\mathbf{X}^{(1)}, \ldots, \mathbf{X}^{(M)}$ from $f(\mathbf{x}; \boldsymbol{\theta}^{(n)})$ and sort it so that $S(\mathbf{X}^{(1)}) \leq \cdots \leq S(\mathbf{X}^{(M)})$. We denote the $j$th component of $\mathbf{X}^{(i)}$ by $X_j^{(i)}$.
\item \textbf{Updating.} Calculate the sample means and standard deviations of the elite sample. That is, calculate
$$
\overline{X}_1 = \frac{1}{\lceil \rho M \rceil}\sum_{i = 1}^{\lceil \rho M \rceil} X_1^{(i)}, \quad \ldots \;, \; \overline{X}_m = \frac{1}{\lceil \rho M \rceil} \sum_{i = 1}^{\lceil \rho M \rceil} X_m^{(i)}
$$
and
$$
S_1 = \sqrt{\frac{1}{\lceil \rho M \rceil - 1} \sum_{i = 1}^{\lceil \rho M \rceil} \left(X_1^{(i)} - \overline{X}_1\right)^2}, \quad \ldots \;, \; S_m = \sqrt{ \frac{1}{\lceil \rho M \rceil - 1} \sum_{i = 1}^{\lceil \rho M \rceil} \left(X_m^{(i)} - \overline{X}_m\right)^2}.
$$
Set $\mu_1^{(n+1)} = \overline{X}_1, \; \ldots, \; \mu_m^{(n+1)} = \overline{X}_m$ and $\sigma_1^{(n+1)} = S_1, \; \ldots, \; \sigma_m^{(n+1)} = S_m$. \newline Set $\boldsymbol{\theta}^{(n+1)} = (\mu_1^{(n+1)}, \sigma_1^{(n+1)}, \ldots, \mu_m^{(n+1)}, \sigma_m^{(n+1)})$.
\item \textbf{Iteration.} If a predetermined stopping condition is met, terminate. Otherwise, set $n = n + 1$ and repeat from step $2$.
\end{enumerate}

\subsubsection{Parameter choice and stopping conditions}

The parameters of the CE algorithm help to improve the speed of convergence and quality of solutions. For example, a good choice of $\boldsymbol{\theta}^{(0)}$ improves the convergence properties of the algorithm. The means, $\mu^{(0)}_1, \ldots, \mu^{(0)}_m$, should be chosen as close as possible to the optimal solution. The initial standard deviations, $\sigma_1^{(0)}, \ldots, \sigma^{(0)}_m$, should be chosen large enough that the algorithm is able to escape local minima but not so large that good configurations are quickly abandoned. 

Ideally, the size of the sample in each step, $M$, should be quite large. However, there are often memory and performance constraints that limit this size. The parameter $\rho$, which controls the size of the elite sample, should be chosen large enough that a representative sample of good solutions are included in the elite sample. However, if $\rho$ is chosen too large, the algorithm will take too long to converge. In general, the bigger $M$ is, the smaller $\rho$ should be.

A standard stopping condition for the algorithm is that it terminates when the cost function does not decrease significantly for a given number of steps.

\subsubsection{Variance injection and dynamic smoothing}

When using the CE approach outlined above, the standard deviations of the normal densities may shrink too quickly. In this case, the CE algorithm can converge to a sub-optimal solution. In order to guard against this, we use variance injection. The basic idea is to occasionally increase the variances of the distributions so that the algorithm can easily escape local minima. Usually, variance is injected when the cost function does not decrease significantly enough over a given number of iterations. The magnitude of this increase can depend on the current value of the cost function. If variance injection is used a number of times without a significant decrease in the cost function, then the algorithm is terminated. 

An alternative to variance injection is to use smoothing when updating $\boldsymbol{\theta}$. The updating step for the means at the $n$th iteration is then of the form
$$
\mu_1^{(n+1)} = \alpha \cdot \overline{X}_1 + (1-\alpha) \cdot \mu_1^{(n)}, \quad \ldots \;, \quad \mu_m^{(n+1)} = \overline{X}_m + (1-\alpha) \cdot \mu_m^{(n)},
$$
and the updating step for the standard deviations is given by
$$
\sigma_1^{(n+1)} = \beta \cdot S_1 + (1-\beta)\cdot \sigma_1^{(n)}, \quad \ldots \;, \quad \sigma_m^{(n+1)} = \beta \cdot S_m + (1-\beta)\cdot \sigma_m^{(n)},
$$
where $\alpha, \beta \in (0, 1]$ (typically $[0.7,1]$). It is also possible to carry out dynamic smoothing, with both $\alpha$ and $\beta$ taken to be functions of the number of steps. In our experience, variance injection is more effective than smoothing. For a detailed discussion of both variance injection and dynamic smoothing, see \cite{Kroese06}. 

\subsection{Solving the LAP}

We solve the LAP by minimizing the approximate discrepancy described in Section \ref{sec:approx}. That is, we solve
\begin{align}
\label{eq:ce_opt}
\{(\mathbf{x}^*_i,r^*_i) \}_{i \in \mathcal{I}} &= \argmin_{\{ (\mathbf{x}_i,r_i) \}_{i \in \mathcal{I}}} \widetilde{\mathcal{D}} \left( \{ (\mathbf{x}_i,r_i) \}_{i \in \mathcal{I}}\right) \nonumber\\
&= \argmin_{\{ (\mathbf{x}_i,r_i) \}_{i \in \mathcal{I}}} \sum_{j=1}^N \, \sum_{k=j+1}^N \widetilde{\mathcal{D}}_{j,k} \left(\mathbf{x}_j, r_j, \mathbf{x}_k, r_k \right),
\end{align}
where $\widetilde{\mathcal{D}}$ is given in \eqref{eq:disc_approx}. Putting this in the terminology of Section \ref{sec:CE_description}, we find the arguments that minimize the cost function $\widetilde{D}$, namely the $N$ generating points, each of which is described by three coordinates and an associated radius. We associate a normal distribution with each of the $m=4N$ values that need to be determined. Thus, for $i = 1, \ldots, N$, the coordinates and radius of the $i$th generating point, $(\mathbf{x}_i, r_i) = (x_i, y_i, z_i, r_i)$, are each described by a normal density. We denote the initial means and standard deviations by $\mu^{(0)}_{x_i}, \mu^{(0)}_{y_i}, \mu^{(0)}_{z_i}, \mu^{(0)}_{r_i}$ and $\sigma^{(0)}_{x_i}, \sigma^{(0)}_{y_i}, \sigma^{(0)}_{z_i}, \sigma^{(0)}_{r_i}$. In order to ensure that the radii are positive, we truncate the corresponding normal densities to the positive real line (this is possible without changing the updating rules, see, e.g., \cite{Rubinstein04,Kroese06}). 

\subsubsection{Variance injection and stopping conditions}
\label{sec:vi_details}

We apply variance injection when the cost function does not decrease significantly over a period of $\tau > 0$ iterations. More precisely, at each iteration of the algorithm, we record $\widetilde{\mathcal{D}}^{(n)}_{\mathsf{min}}$, the minimal value of the approximate discrepancies calculated from the sample in that step. If, at the $n$th step,
$$
\left|\frac{\widetilde{\mathcal{D}}^{(n)}_{\mathsf{min}} - \max_{t \in \{n - \tau, \ldots, n - 1\}} \widetilde{\mathcal{D}}^{(t)}_{\mathsf{min}}}{\widetilde{\mathcal{D}}^{(n)}_{\mathsf{min}}}\right| < \delta_{\mathsf{inject}},
$$
we perform variance injection. Because many generating points may already be close to their optimal positions and sizes, variance injection is carried out locally with a magnitude controlled by a parameter $\kappa > 0$. This is done by calculating the local cost of each cell and increasing the variance of the associated densities accordingly. We calculate the average cost of the $i$th cell by
\begin{align*}
\overline{\mathcal{D}}_i &= \frac{1}{\sum_{j=1}^N \# T_{i,j}} \sum_{j=1}^N \, \widetilde{\mathcal{D}}_{i,j} \left((\mu^{(n)}_{x_i}, \mu^{(n)}_{y_i}, \mu^{(n)}_{z_i}), \mu^{(n)}_{r_i}, (\mu^{(n)}_{x_j}, \mu^{(n)}_{y_j}, \mu^{(n)}_{z_j}), \mu^{(n)}_{r_j}\right).
\end{align*}
The local cost of a cell, $\mathcal{D}^\star_i$, is defined to be the maximum of its own average cost and the average cost of its adjacent cells. That is,
\begin{equation}\label{eqn:local-cell-cost}
\mathcal{D}^\star_i = \max\left\{ \overline{\mathcal{D}}_k : k=i \text{ or } k \text{ such that } N_{i,k}^\star \neq \emptyset \right\}, \quad i=1,\ldots,N \,.
\end{equation}
The variance injection is performed by setting
\begin{align*}
&\sigma_{x_i}^{(n)} = \sigma_{x_i}^{(n)} + \kappa \cdot \sqrt{\mathcal{D}^\star_i}, \quad \sigma_{y_i}^{(n)} = \sigma_{y_i}^{(n)} + \kappa \cdot \sqrt{\mathcal{D}^\star_i}, \\
&\sigma_{z_i}^{(n)} = \sigma_{z_i}^{(n)} + \kappa \cdot \sqrt{\mathcal{D}^\star_i}, \quad \sigma_{r_i}^{(n)} = \sigma_{r_i}^{(n)} + \kappa \cdot \sqrt{\mathcal{D}^\star_i},
\end{align*}
for  $i = 1, \ldots, N$.

If, at any stage, the benefit of variance injection becomes negligible, we stop performing it. More precisely, if the current minimum cost, ${\mathcal{D}}^{(n)}_{\mathsf{min}}$ divided by the minimum cost immediately prior to the last variance injection is larger than $\gamma \in (0,1)$, we no longer carry out variance injection.

The algorithm is terminated when 
$$
\left|\frac{\widetilde{\mathcal{D}}^{(n)}_{\mathsf{min}} - \max_{t \in \{n - \tau, \ldots, n - 1\}} \widetilde{\mathcal{D}}^{(t)}_{\mathsf{min}}}{\widetilde{\mathcal{D}}^{(n)}_{\mathsf{min}}}\right| < \delta_{\mathsf{terminate}},
$$
where $\delta_{\mathsf{terminate}} < \delta_{\mathsf{inject}}$.

\subsubsection{Initial configuration}\label{sec:optimization:initial}

Reasonably good initial tessellations can be obtained directly from the tomographic data \cite{Lyckegaard10}. The centroids and equivalent radii of the grains in the tomographic data are used for the coordinates and radii of the generating points. The centroids, $\{(c_{x_i}, c_{y_i}, c_{z_i})\}_{i \in \mathcal{I}}$, are given by
$$
c_{x_i} = \sum_{(x,y,z) \in R_i} \frac{x}{\# R_i}, \quad c_{y_i} = \sum_{(x,y,z) \in R_i} \frac{y}{\# R_i}, \quad c_{z_i} = \sum_{(x,y,z) \in R_i} \frac{z}{\# R_i},
$$
for $i=1,\ldots,N$. The equivalent radii are given by
$$
\widehat{r}_i = \sqrt[3]{\frac{3}{4\pi} \# R_i}
$$
for $i=1,\ldots,N$. The initial means of the densities are then given by $\mu_{x_i}^{(0)} = c_{x_i}$, $\mu_{y_i}^{(0)} = c_{y_i}$, $\mu_{z_i}^{(0)} = c_{z_i}$ and $\mu_{r_i}^{(0)} = \widehat{r}_i$ for $i = 1, \ldots, N$.

The initial standard deviations are chosen proportional to the local cost of the cells in this approximation. The local costs are calculated as in Section~\ref{sec:vi_details}. Thus,
$$
\sigma_{x_i}^{(0)} = \sigma_{y_i}^{(0)} = \sigma_{z_i}^{(0)} = \sigma_{r_i}^{(0)} = \sqrt{\mathcal{D}^\star_i},\quad 
$$
for $i = 1, \ldots, N$.

\subsubsection{Choice of control parameters for the CE method}

Our investigations showed that a parameter choice of $M = 4000$, $\rho = 0.05$, $\delta_{\text{inject}} = 0.05$, $\delta_{\text{terminate}} = 0.01$, $\tau = 10$, $\gamma = 0.9$ and $\kappa = 0.25$ works well for a large number of different data sets. The elite set then consists of $\lceil \rho M \rceil = 200$ samples, which results in a sufficiently large sample of good solutions. Note that other parameter values may improve the speed of convergence (or, in the case of the parameters controlling variance injection and termination, improve the quality of the approximation). However, the basic performance of the CE algorithm is relatively robust to parameter choice. That is, the convergence behavior and quality of approximations are good for a wide range of parameters. 

\subsubsection{Edge cells}

The approximations obtained using the CE algorithm have unbounded cells at the edge of the observation window. This is because the corresponding grains in the tomographic data are not delimited by other grains. The unbounded cells are then intersected with the observation window, $W$. Note that, in most cases, edge grains are of little scientific interest and are not considered when analyzing the data. In practice, edge grains may be only partially observed or may not be representative (e.g., when stuying the dynamics of grains undergoing grain coarsening). In this paper, we explicitly ignore grains and their approximating cells if they lie at the edge of the observation window.

\section{Experimental results}\label{sec:results}

In this section, we present the results of a number of numerical experiments which we carried out in order to demonstrate the effectiveness of the CE approach. We consider three distinct data sets. Two of these data sets are produced using stochastic models. These `artificial' data sets allow us to investigate how well our method is able to reconstruct tessellations when we are certain the underlying tessellation is, indeed, Laguerre. The first artificial data set is produced using a stochastic model that describes a polycrystalline material undergoing grain coarsening. The second artificial data set is, by design, much more pathological. It exhibits large variation in the size and shape of its grains, as well as the number of neighbors each grain has. This data set allows us to explore the effectiveness of our approach in an extreme setting. In particular, it provides an example of a setting in which the standard choice of initial conditions is far from optimal. Finally, we consider an empirical data set: tomographic data obtained from a sample of Al-5 wt\% Cu. We demonstrate that our method is able to produce an excellent approximation to this data set using a Laguerre tessellation.

\subsection{Artificial data}\label{sec:results:artificial}

\begin{figure}
\begin{center}
\subfigure[Data set ``artificial-PCM''.]{\tightfbox{\includegraphics[width=0.45\textwidth,trim=5cm 5cm 5cm 5cm,clip]{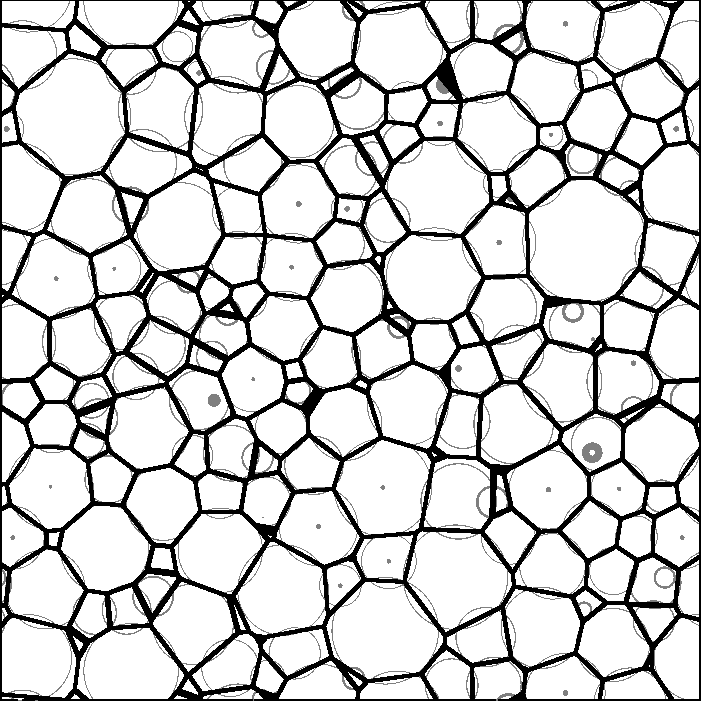}}}
\subfigure[Data set ``artificial-PLT''.]{\tightfbox{\includegraphics[width=0.45\textwidth,trim=5cm 5cm 5cm 5cm,clip]{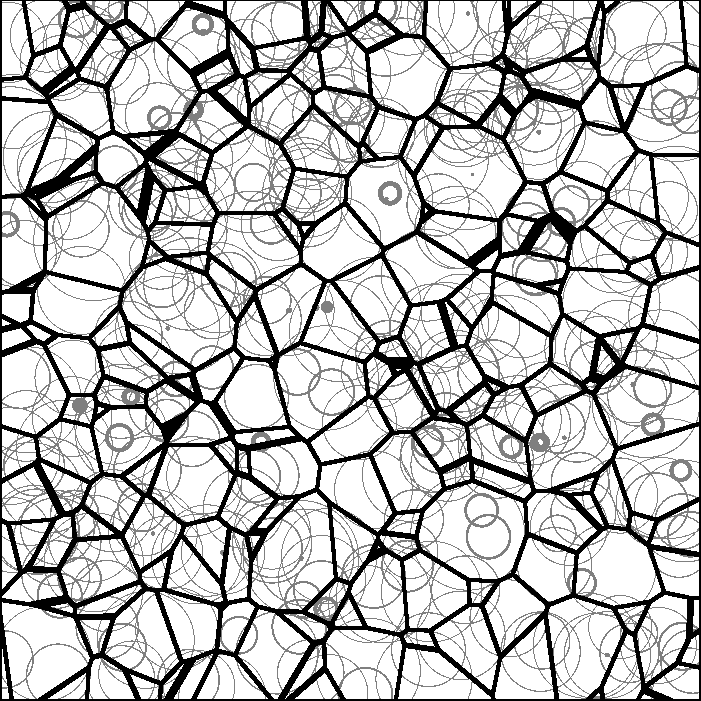}}}
\caption{Partial cross-sections of artificial data sets. Spheres used as generators are drawn in gray (dots for sphere centers, gray circles for sphere surfaces). Radii are very large in the PLT case, therefore many surface voxels of spheres from other slices are visible.}
\label{fig:artificial:cross-sections}
\end{center}
\end{figure}

The artificial data sets are produced using two distinct stochastic models. The polycrystalline model (PCM), developed in \cite{Spettl15b}, describes a polycrystalline material undergoing grain coarsening. The second data set is produced using a randomly marked Poisson process; see \cite{Illian08}. The resulting tessellation is known as a Poisson-Laguerre tessellation (PLT); see \cite{Lautensack07}. In the first case, there is a strong correlation between the relative positions of the generating points and their weights. In the second example, the weights are independent of the positions of the generating points. As a result, the second data set is much less regular than the first data set. Figure~\ref{fig:artificial:cross-sections} shows 2D cross sections of the data sets, together with their generating points.

\subsubsection{Model descriptions}

In the PCM model, a random Laguerre tessellation is produced using a set of spheres that may overlap slightly. The centers of the spheres give the locations of the generating points and the radii define the weights. The spheres have hard-cores which may not overlap. This was shown to be a suitable model for polycrystalline materials in \cite{Spettl15b}. Both the density of the sphere packing and the radii of the spheres influence the sizes and shapes of the cells in the resulting tessellation. In particular, when highly dense packings are used, very ``spherical'' cells are generated with a narrow coordination number distribution. We generated a sample from this model in a bounded window, $W=[0,700]^3$ using parameters that were fitted to a sample of Al-5 wt\% Cu annealed for 200 minutes. The fitting procedure is described in \cite{Spettl15b}. Note that the parameters we use were obtained by fitting the model to the empirical data described in Section~\ref{sec:results:experimental}. Having produced a sample, we removed cells that would correspond to a very small number of voxels after discretization (i.e., cells having a smaller volume than a ball with radius of 4 voxels). The final result is a tessellation consisting of roughly 2500 grains (approximately the same number as in the experimental data).

The PLT model is generated by simulating a homogeneous Poisson process with some intensity $\lambda$ in a bounded window $W$. The points are independently marked by gamma-distributed random variables with some shape parameter $\alpha$ and rate parameter $\beta$. Note that, unlike the PCM model, it is possible that some generating points will not produce cells in the PLT model. We generated a sample from this model in $W = [0, 700]^3$ with intensity $\lambda = 7.5\times10^{-6}$ (so that the expected number of points is roughly 2500), where we used parameters $\alpha = 50$ and $\beta = 1.125$ for the mark distribution.

\subsubsection{Laguerre approximation}

\begin{figure}[htb]
\begin{center}
\subfigure[PCM data.]{\tightfbox{\includegraphics[width=0.45\textwidth,trim=5cm 5cm 5cm 5cm,clip]{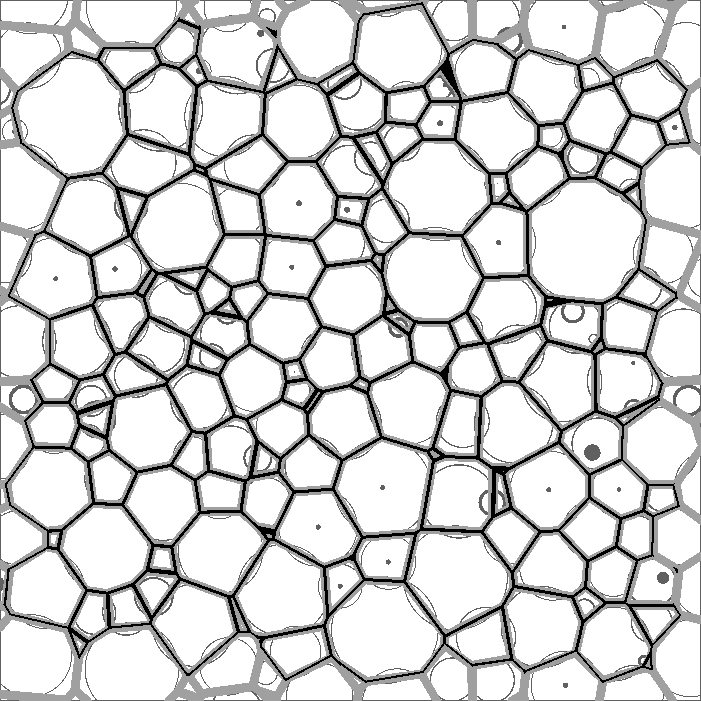}}}
\subfigure[PLT data.]{\tightfbox{\includegraphics[width=0.45\textwidth,trim=5cm 5cm 5cm 5cm,clip]{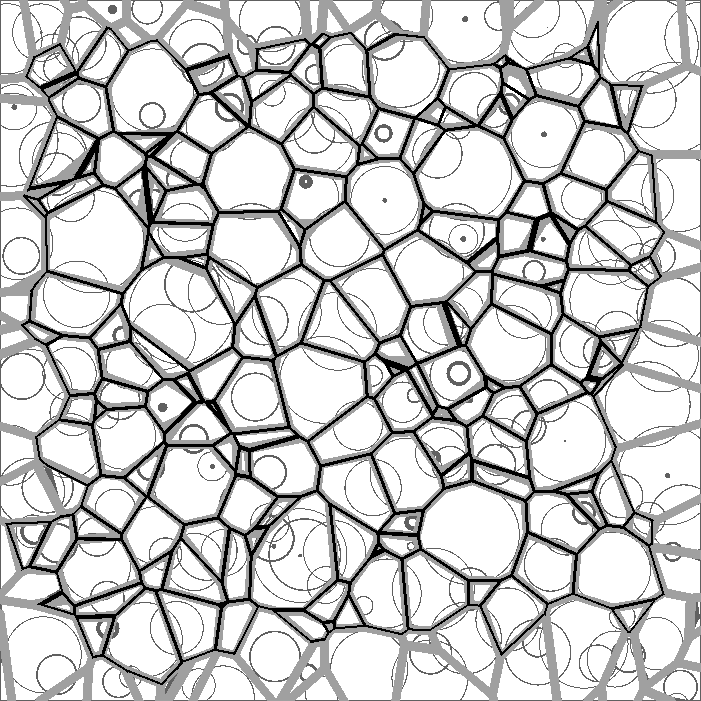}}}
\caption{Partial cross-sections of artificial data: original cell boundaries (gray) superimposed with cell boundaries of our Laguerre approximation (black), with centers and radii of detected generators drawn in small gray dots and thin gray circles.}
\label{fig:artificial:cross-sections-overlay}
\end{center}
\end{figure}

The realizations of the PCM and PLT models were both discretized on a voxel grid, resulting in two $700 \times 700 \times 700$ images. These images are of the form described in Section~\ref{sec:data}. The CE method was then used to solve the LAP. The results are illustrated in Figure~\ref{fig:artificial:cross-sections-overlay}, where 2D cross-sections of the original tessellations are shown with the approximating tessellations superimposed. 

Using multi-threading on a standard quad-core processor (Intel Core i5-3570K), the computing time was roughly 3 hours for the PCM data and 4 hours for the PLT data. For smaller test sets, with 500 grains each, the time required was roughly 20 to 30 minutes. The memory requirements when fitting the full data sets were minimal (especially in comparison to approaches where it is necessary to store the complete voxelized data). Basically, it is necessary to store the test points and all the generators from one iteration of the CE method. The cost of storing additional variables, such as the means and standard deviations of the densities describing the generators and the cost values of the test points are negligible. For the PCM data, the number of test points used was approximately 140\,000. Approximately 160\,000 points were used for the PLT data. Together with $N \cdot M \approx 2500 \cdot 4000$ generators in one iteration, this sums up to less than 200~MB of RAM when using 32-bit floating point coordinates.

\begin{figure}[htb]
\begin{center}
\subfigure[PCM data: cost values in CE iterations.]{\includegraphics[width=0.45\textwidth,trim=0cm 1cm 1cm 2cm,clip]{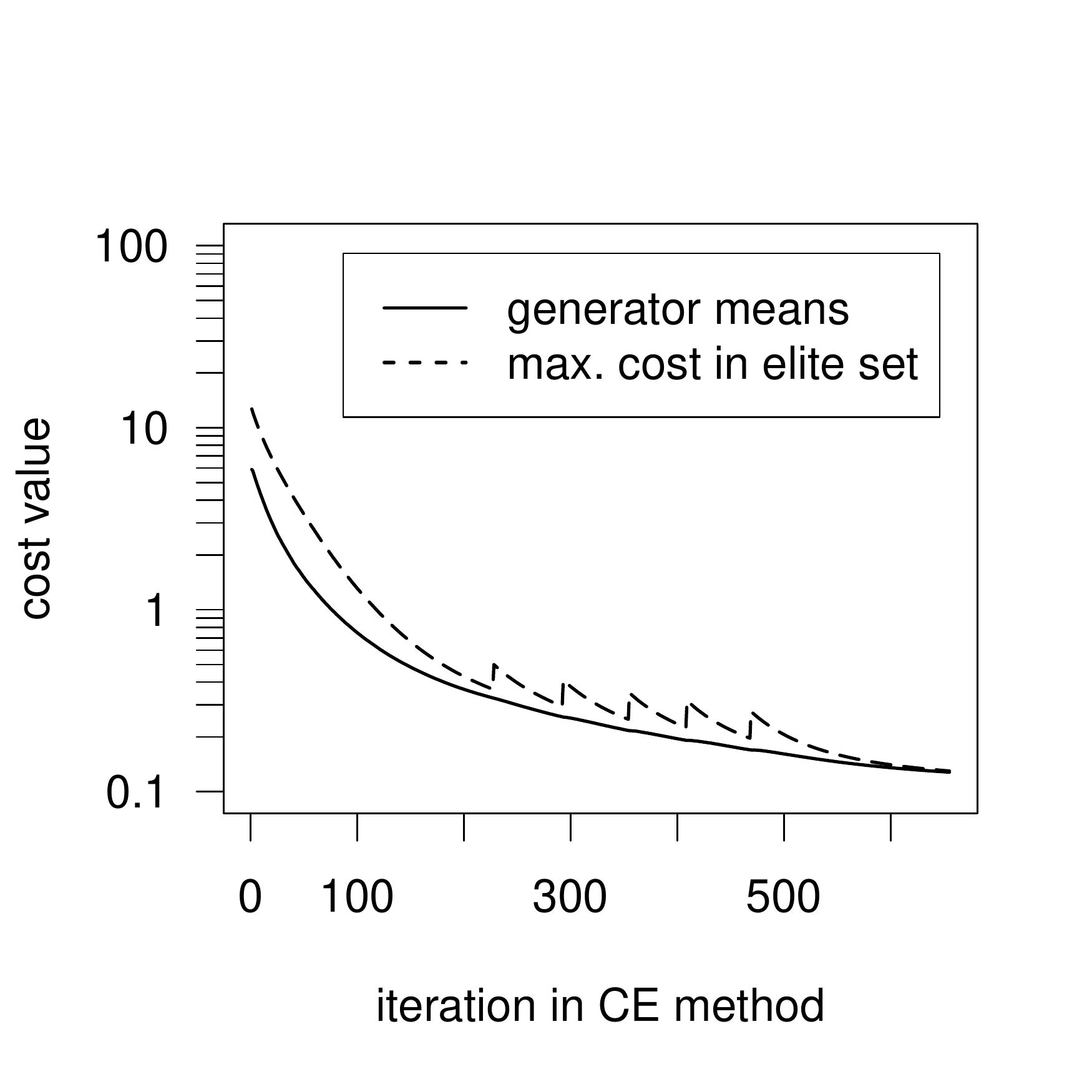}}
\subfigure[PCM data: maximum of standard deviations in CE iterations.]{\includegraphics[width=0.45\textwidth,trim=0cm 1cm 1cm 2cm,clip]{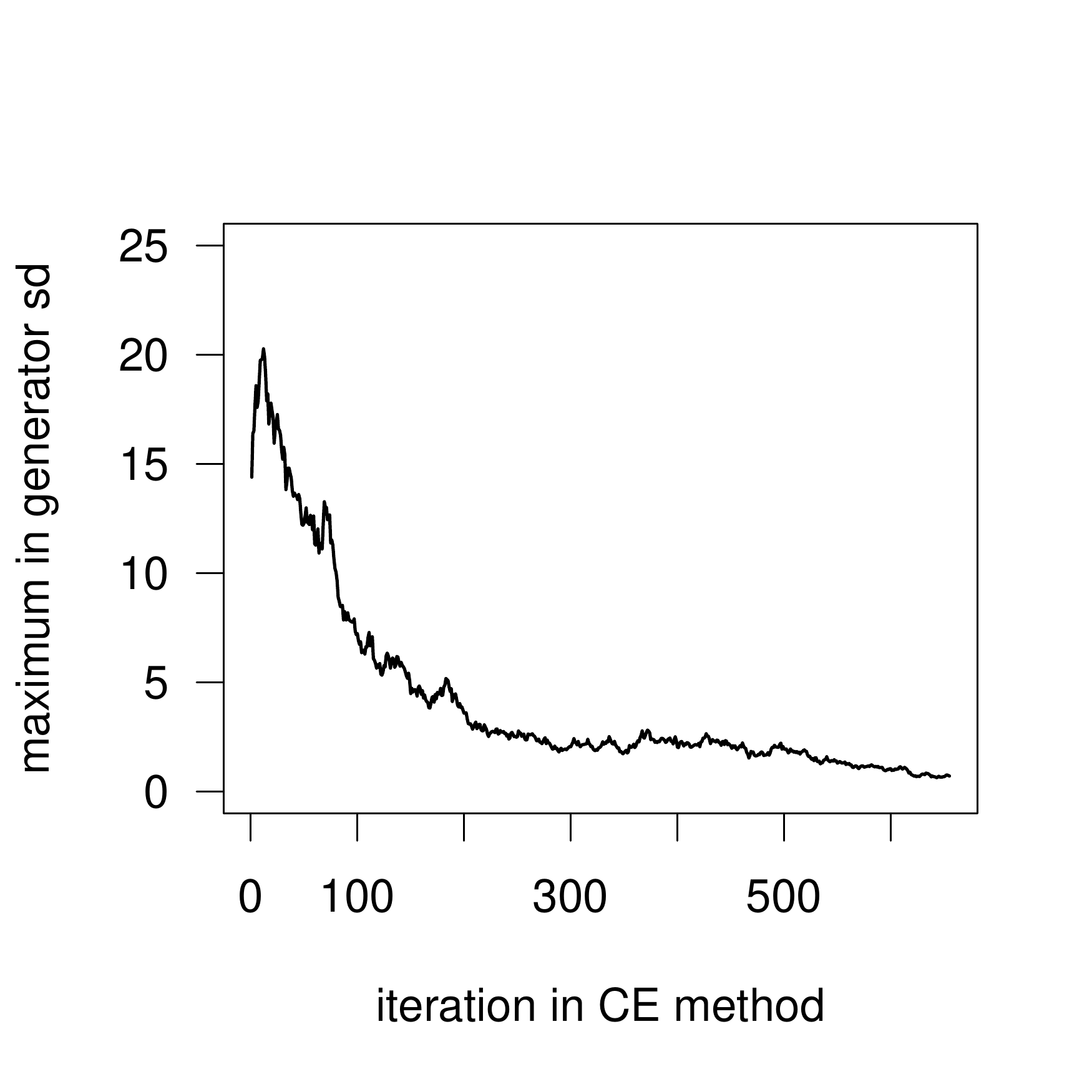}}
\subfigure[PLT data: cost values in CE iterations.]{\includegraphics[width=0.45\textwidth,trim=0cm 1cm 1cm 2cm,clip]{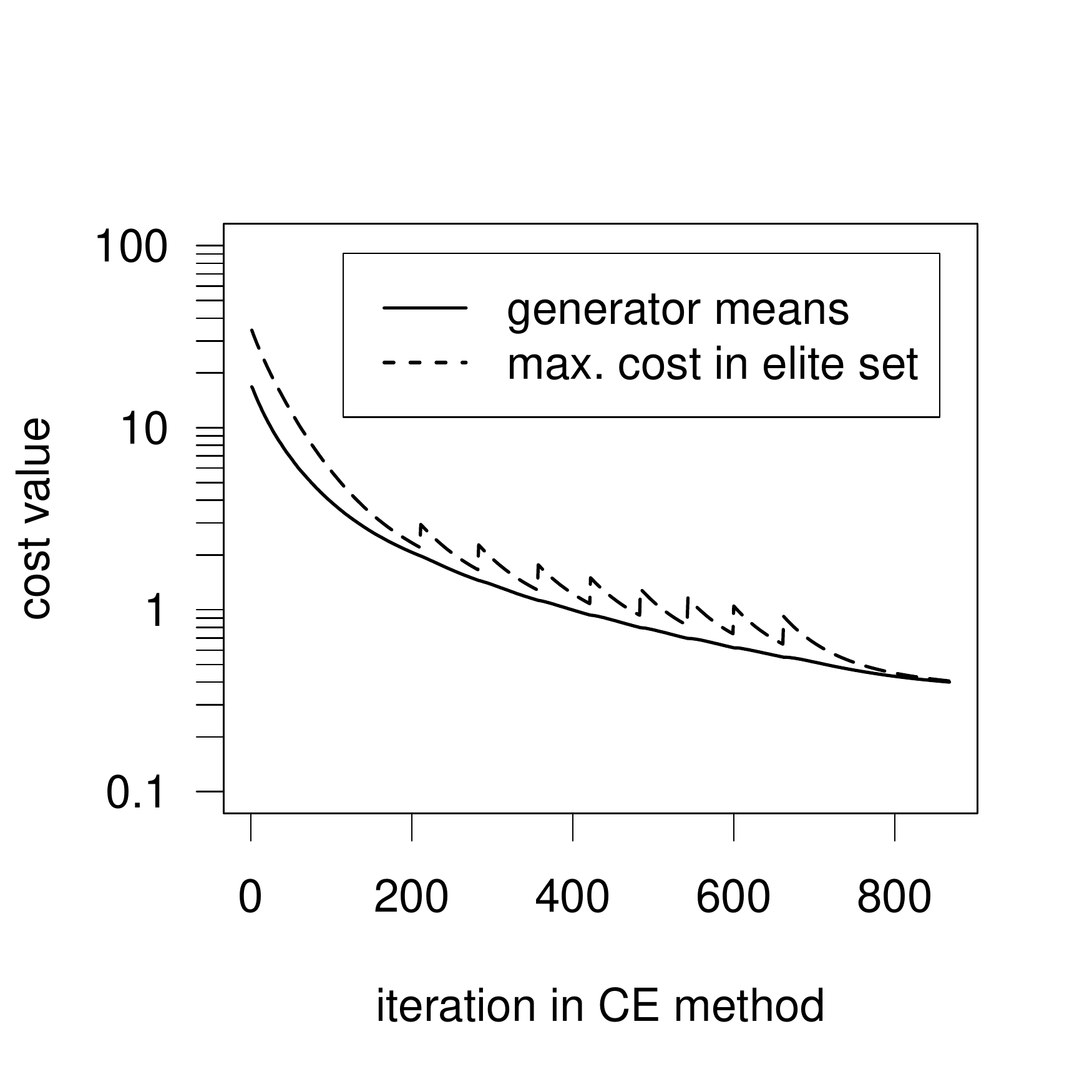}}
\subfigure[PLT data: maximum of standard deviations in CE iterations.]{\includegraphics[width=0.45\textwidth,trim=0cm 1cm 1cm 2cm,clip]{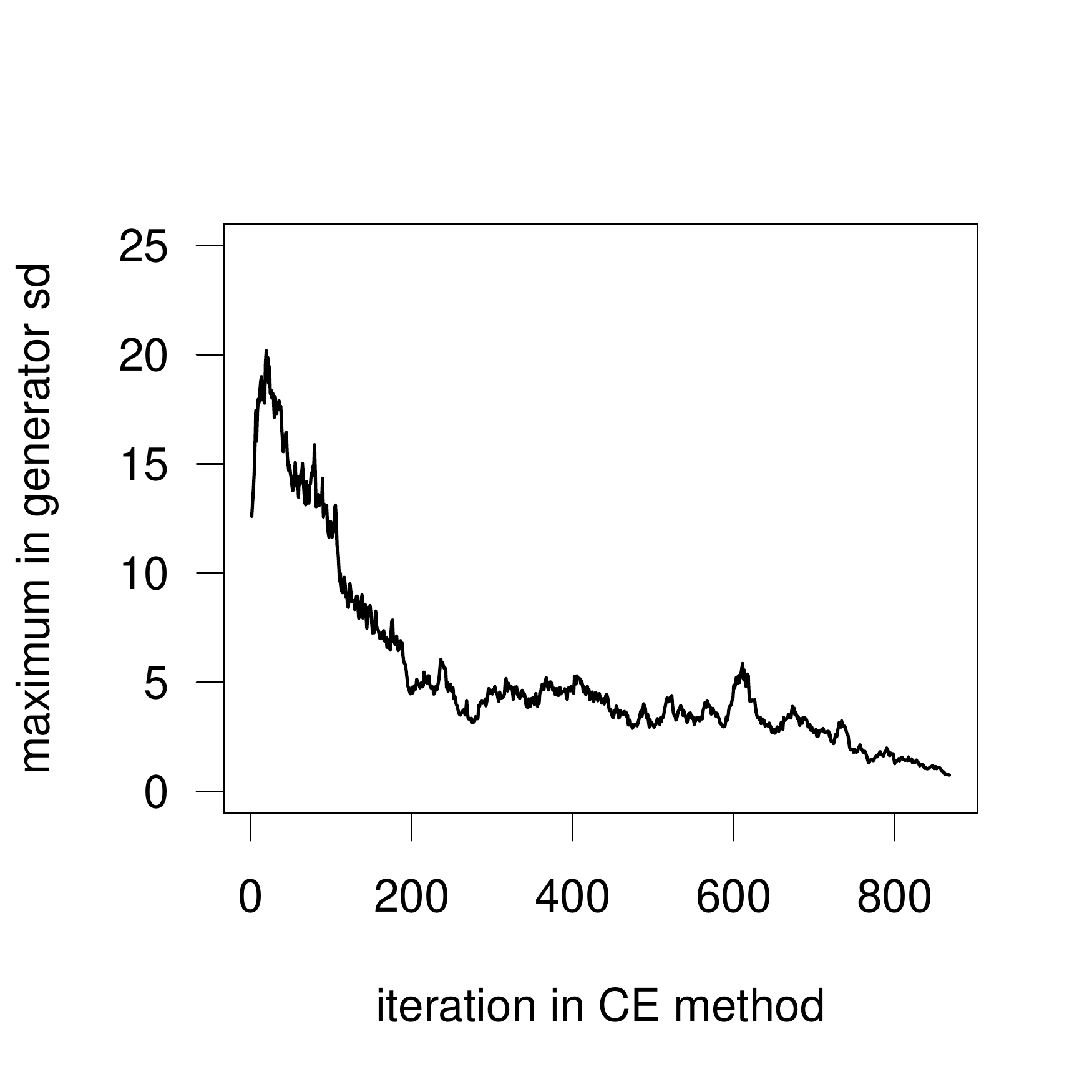}}
\caption{CE method for artificial data: convergence of the cost values $\widetilde{\mathcal{D}}$ (Eq.~\eqref{eq:disc_approx}) and standard deviations.}
\label{fig:artificial:convergence}
\end{center}
\end{figure}

Figure~\ref{fig:artificial:convergence} illustrates the convergence behavior of the CE algorithm. When approximating the PCM data, roughly 650 iterations of the algorithm were required. In the PLT case, about 900 iterations were required. It is not surprising that the CE algorithm took longer to terminate in the PLT case. As will be seen in Section~\ref{sec:results:artificial:evaluation} the initial conditions in this setting are much further from the optimal solution. 

\subsubsection{Results}\label{sec:results:artificial:evaluation}

In order to evaluate the quality of our approximations, we compare them to approximations obtained using the heuristic approach presented in \cite{Lyckegaard10} and the orthogonal regression method proposed in \cite{Spettl14}. Note that, although the orthogonal regression approach is able to achieve quite good approximations of the cells, it does not result in a parametrized tessellation. Table~\ref{tab:artificial:evaluation} shows an evaluation of the approximations with respect to a volume-based discrepancy measure: the number of voxels that are correctly labeled. Our method results in an almost perfect approximation of the PCM data. The heuristic approach considered in \cite{Lyckegaard10} also works very well. However, it is not able to reproduce the neighbor structure as successfully as our approach, as evidenced by rows 2 through 5 of Table~\ref{tab:artificial:evaluation}. Our method is slightly less successful at approximating the PLT data but, in this case, considerably outperforms the heuristics of \cite{Lyckegaard10}. 

\begin{table}[htb]
\tbl{Evaluation of the artificial data approximations: H denotes the heuristic approach of \cite{Lyckegaard10}; CE denotes the CE method considered in the present paper; OR denotes orthogonal regression proposed in \cite{Spettl14}.}
{
\begin{tabular}{lccccccc} \toprule
                                            & \multicolumn{3}{c}{PCM}  & & \multicolumn{3}{c}{PLT}  \\
                                            &   H    &   CE   &   OR   & &    H   &  CE    & OR     \\ \colrule
correctly labeled voxels [\%]               & $97.3$ & $99.1$ & $99.8$ & & $87.8$ & $96.2$ & $99.8$ \\
cells with all neighbors correct [\%]       & $75.6$ & $88.5$ & $78.9$ & & $18.6$ & $45.1$ & $61.3$ \\
cells with $\leq1$ incorrect neighbors [\%] & $96.3$ & $99.0$ & $97.3$ & & $46.7$ & $77.3$ & $90.4$ \\
cells with $\leq2$ incorrect neighbors [\%] & $99.6$ & $99.8$ & $99.7$ & & $72.2$ & $92.1$ & $98.2$ \\
mean number of erroneous neighbors/grain    & $0.28$ & $0.13$ & $0.24$ & & $1.83$ & $0.90$ & $0.51$ \\
average displacement of centroids [voxels]  & $0.56$ & $0.24$ & $0.09$ & & $2.56$ & $1.33$ & $0.06$ \\ \botrule
\end{tabular}
}
\label{tab:artificial:evaluation}
\end{table}

The parameters used by the heuristic approach of \cite{Lyckegaard10} are very close to the initial conditions of the CE algorithm. Thus, from the difference in performance, it seems that the initial conditions are quite far from the optimal parameters. In order to find better configurations, the CE algorithm has to be able to escape local minima around these initial conditions. This is made clear by the fact that increasing the number of times variance injection is used (as well as the number of iterations), we are able to further improve the results for the CE algorithm. Namely, by manually increasing the number of variance injections to 20 (instead of the 8 used in the standard approach), we obtained an approximation which correctly labeled 97.3\% of the voxels (instead of 96.2\%).
This significant improvement indicates that local minima are present in which the CE algorithm is becoming trapped (because variance injection works by helping the algorithm escape local minima). This, in turn, implies that there are local minima near to the initial conditions of the CE algorithm that are sufficiently deep to trap it.
A direct implication of these results is that algorithms that converge to nearby local minima (such as gradient-descent) may not lead to good approximations when using a standard choice of initial conditions. Furthermore, it is not clear that it is always straightforward to find good initial conditions. For example, the initial generating points are usually chosen to lie inside their cells. But, using the methods presented in \cite{Duan14} and the exact description of the PLT tessellation, it was not possible to find a solution to the LIP subject to the restriction that generating points were contained in their cells. If a method such as gradient-descent starts with all of the points lying inside their cells, it is unclear that it will be able to obtain a solution where some generating points lie far outside their cells.

\begin{figure}[htb]
\begin{center}
\subfigure[PCM data.]{\includegraphics[width=0.45\textwidth,trim=0cm 1cm 1cm 2cm,clip]{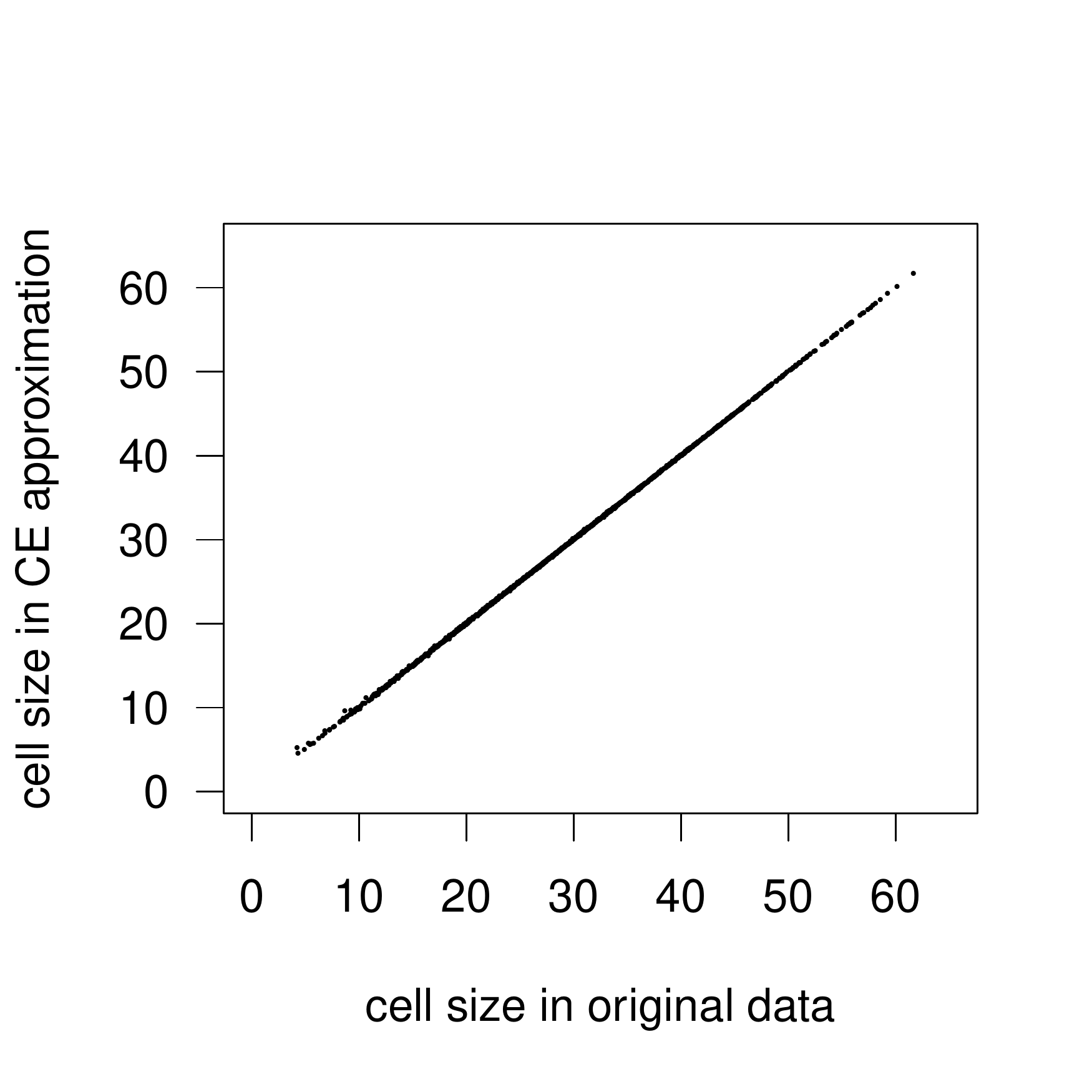}}
\subfigure[PLT data.]{\includegraphics[width=0.45\textwidth,trim=0cm 1cm 1cm 2cm,clip]{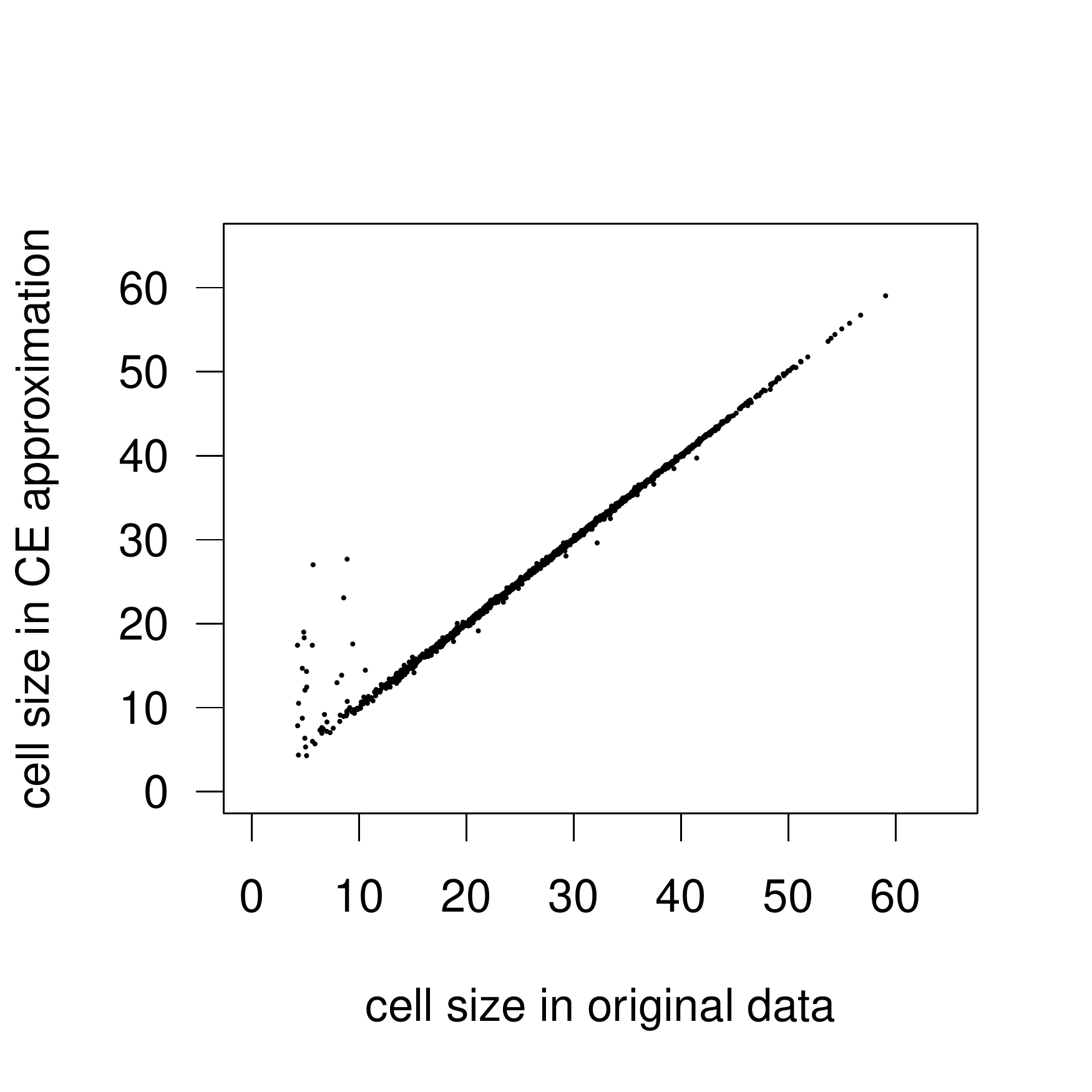}}
\caption{Scatter plot of cell sizes given as radii of volume-equivalent spheres obtained from the original data and cells extracted by the CE method.}
\label{fig:artificial:cell-volumes}
\end{center}
\end{figure}

Figure~\ref{fig:artificial:cell-volumes} shows scatter plots of the original cell sizes vs. the cell sizes in the CE approximations. The quality of the CE approximation of the PCM data is immediately apparent. It is also clear that the difficulties in fitting the PLT data lie primarily in the small cells. 

\subsection{Experimental data}\label{sec:results:experimental}

\begin{figure}[htb]
\begin{center}
\subfigure[Tomographic grayscale data.]{\tightfbox{\includegraphics[height=0.45\textwidth,trim=6cm 6cm 6cm 6cm,clip]{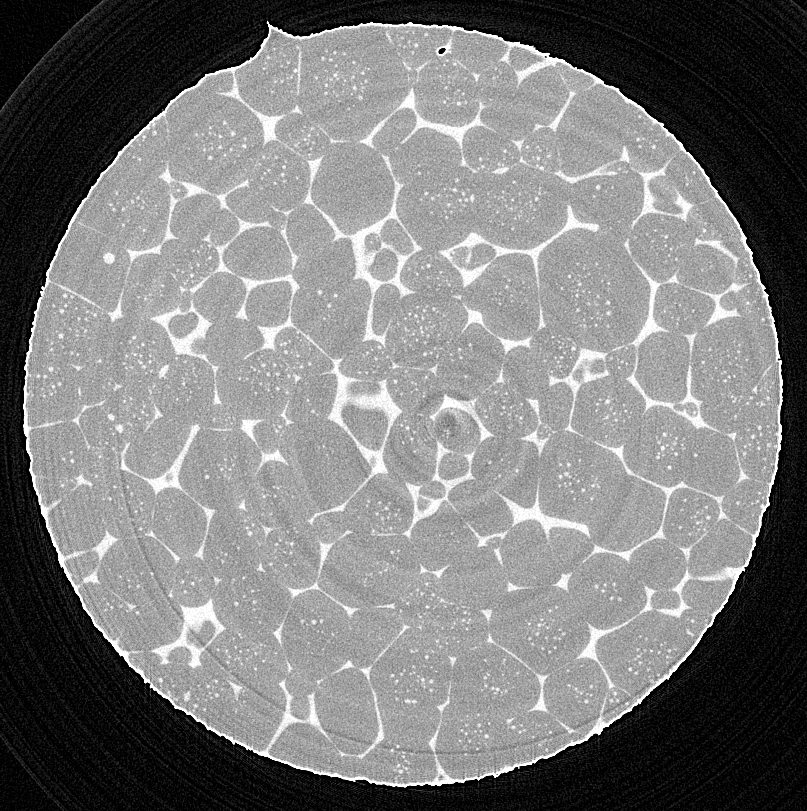}}}
\subfigure[Grain boundaries obtained by watershed segmentation.]{\tightfbox{\includegraphics[height=0.45\textwidth,trim=6cm 6cm 6cm 6cm,clip]{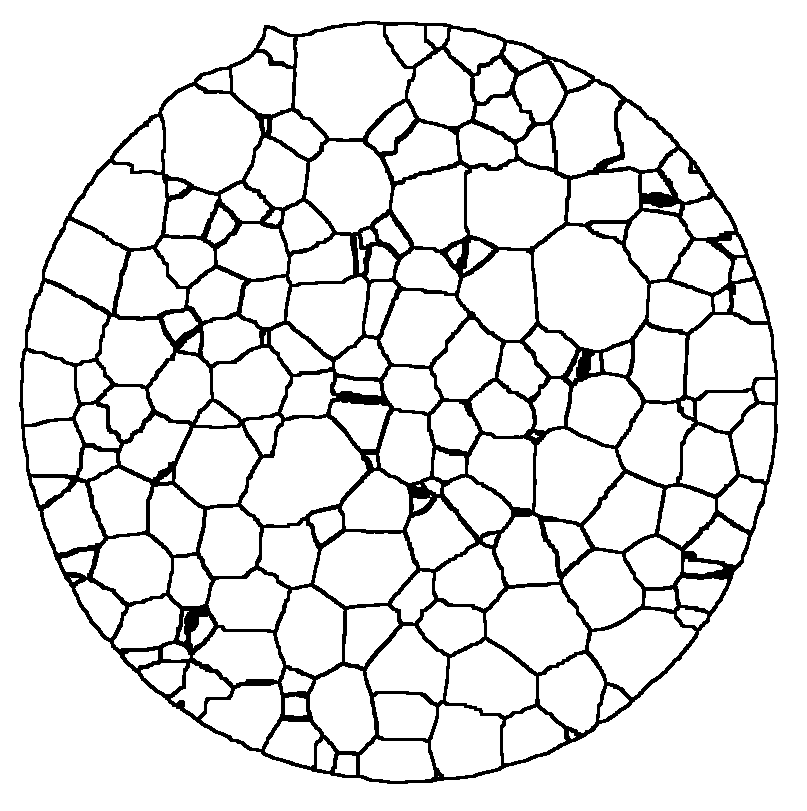}}}
\caption{Partial cross-section of experimental data.}
\label{fig:experimental:cross-section}
\end{center}
\end{figure}

The experimental data we consider was used in \cite{Spettl15b} to develop the PCM model discussed in Section \ref{sec:results:artificial}. An Al-5 wt\% Cu sample (cylindrical with $8.5$\,mm length and $4$\,mm diameter) was heated to the semisolid state at a temperature of $592\,^{\circ}{\rm C}$, at which point coarsening processes were measured \emph{in situ} using synchrotron X-ray tomographic imaging. For the present paper, we consider only data obtained after an annealing time of 200 minutes (which corresponds to the first annealing step in \cite{Spettl15b}). It consists of approximately 2500 grains, which is quite a large data set. A 2D cross-section of the data is shown in Figure \ref{fig:experimental:cross-section}. The experimental data set was segmented using the watershed transformation, resulting in a labeled image of grain regions, which are separated by a watershed layer of one-voxel thickness. Further details regarding sample preparation, imaging and segmentation can be found in \cite{Spettl15b}.

\subsubsection{Laguerre approximation}

\begin{figure}[htb]
\begin{center}
\tightfbox{\includegraphics[height=0.45\textwidth,trim=6cm 6cm 6cm 6cm,clip]{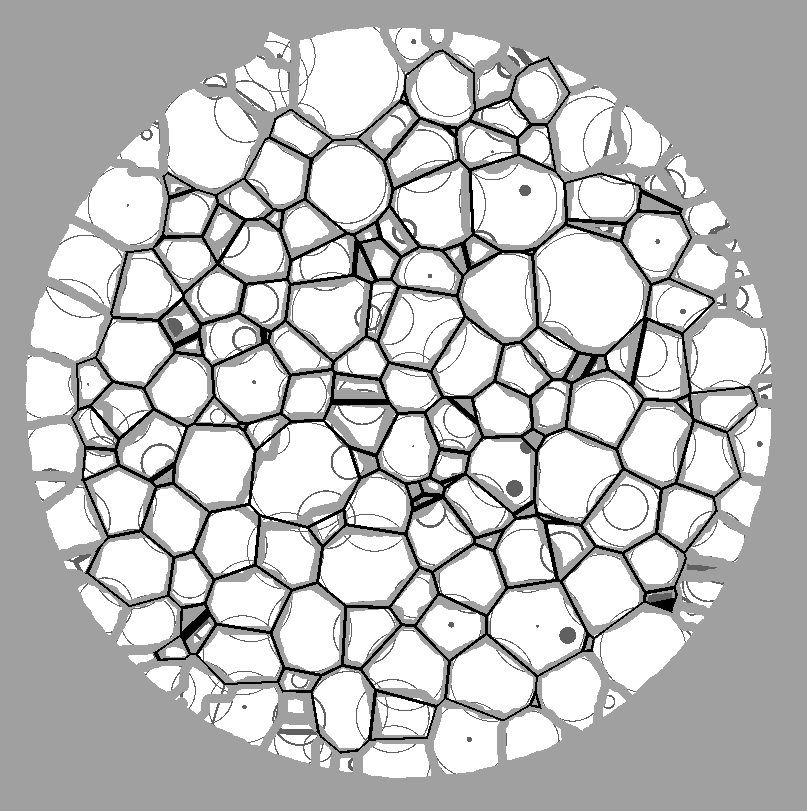}}
\caption{Partial cross-section of experimental data: original boundaries of grain regions (gray) superimposed with cell boundaries of our Laguerre approximation (black), with centers and radii of detected generators drawn in small gray dots and thin gray circles.}
\label{fig:experimental:cross-section-overlay}
\end{center}
\end{figure}

The CE method was applied to solve the LAP for the experimental data set. A cross-section of the resulting Laguerre approximation is given in Figure~\ref{fig:experimental:cross-section-overlay}.

\begin{figure}[htb]
\begin{center}
\subfigure[Cost values for experimental data in CE \mbox{iterations}.]{\includegraphics[width=0.45\textwidth,trim=0cm 1cm 1cm 2cm,clip]{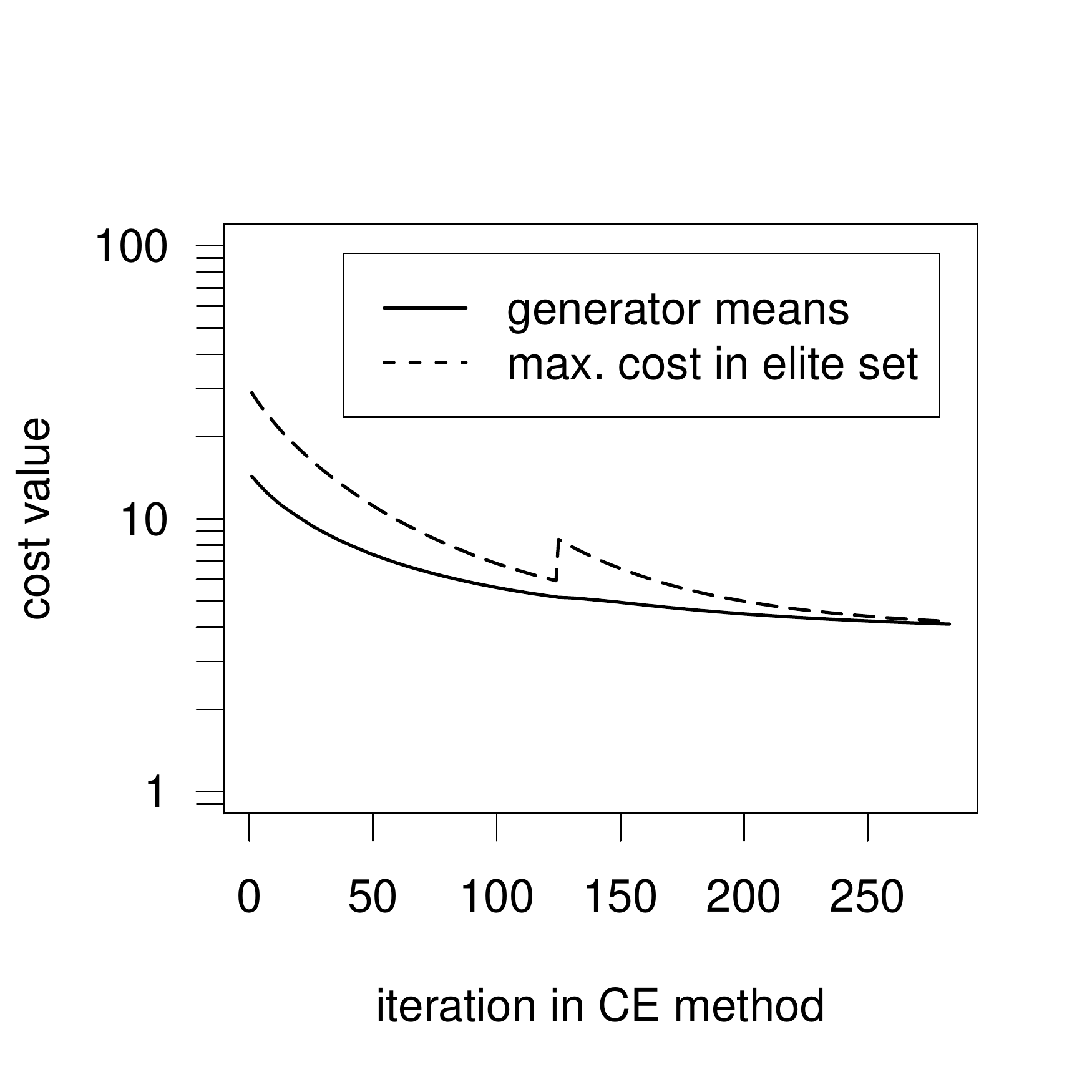}}
\subfigure[Maximum of standard deviations for \mbox{experimental} data in CE iterations.]{\includegraphics[width=0.45\textwidth,trim=0cm 1cm 1cm 2cm,clip]{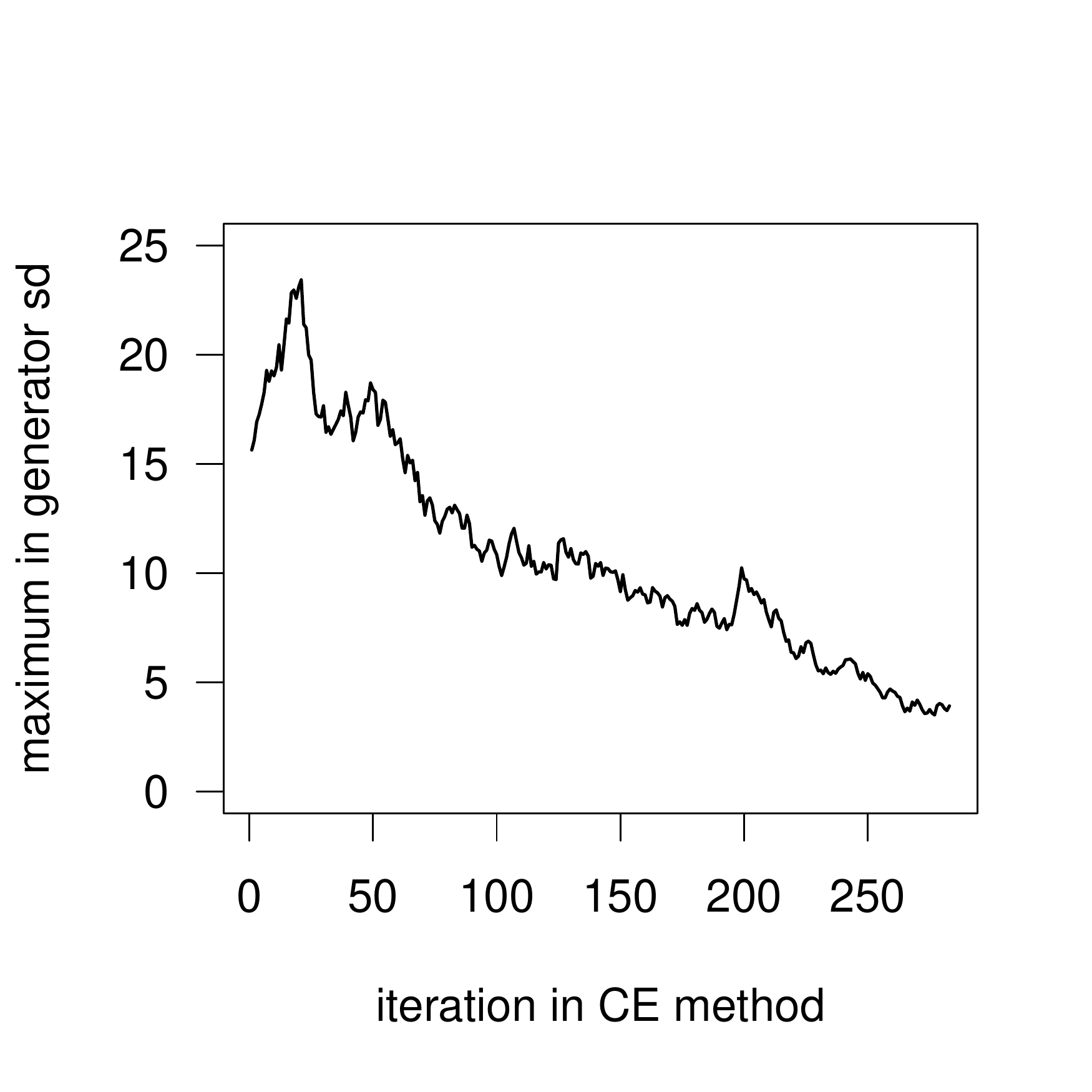}}
\caption{CE method for experimental data: convergence of the cost values $\widetilde{\mathcal{D}}$ (Eq.~\eqref{eq:disc_approx}) and standard deviations.}
\label{fig:experimental:convergence}
\end{center}
\end{figure}

Fitting the approximation took approximately 70 minutes on the same Intel Core i5-3570K quad-core processor used to fit the artificial data.  The number of test points extracted from the image data was roughly 150\,000, which is almost the same as for the artificial data sets. The convergence behavior of the CE method is illustrated in Figure~\ref{fig:experimental:convergence}. Using the same parameters, only 283 iterations were necessary, which is substantially fewer than the number required for the artificial data. This is mainly due to the fact that only one variance injection was required.

\subsubsection{Results}

As with the artificial data, we compare our results with the heuristic approach proposed in \cite{Lyckegaard10} and the orthogonal regression approach introduced in \cite{Spettl14}. As mentioned above, the orthogonal regression approach reconstructs the individual cells quite well but does not result in a tessellation. The results are summarized in Table~\ref{tab:experimental:evaluation}. The CE method correctly assigns $91.4$\,\% of the voxels. In contrast, the heuristic approach correctly assigns only $82.9$\,\% of the voxels. The orthogonal regression yields $96.3$\,\%.

\begin{table}[htb]
\tbl{Evaluation of the experimental data approximations: H denotes the heuristic approach of \cite{Lyckegaard10}; CE denotes the CE method considered in the present paper; OR denotes orthogonal regression proposed in \cite{Spettl14}.}
{
\begin{tabular}{lccc} \toprule
                                            &    H   &   CE   &   OR   \\ \colrule
correctly labeled voxels [\%]               & $82.9$ & $91.4$ & $96.3$ \\
cells with all neighbors correct [\%]       & $20.1$ & $38.4$ & $29.3$ \\
cells with $\leq1$ incorrect neighbors [\%] & $48.4$ & $73.1$ & $50.2$ \\
cells with $\leq2$ incorrect neighbors [\%] & $65.6$ & $90.6$ & $62.6$ \\
mean number of erroneous neighbors/grain    & $2.29$ & $1.04$ & $2.37$ \\
average displacement of centroids [voxels]  & $3.40$ & $1.78$ & $0.23$ \\ \botrule
\end{tabular}
}
\label{tab:experimental:evaluation}
\end{table}

Note that the CE method significantly outperforms both the heuristic method and orthogonal regression when describing the neighborhood structure of the grains, although it does not seem possible for a tessellation containing only convex cells to accurately capture the full neighborhood structure. This is at least partially due to segmentation issues and contacts in the image data with very small areas. In contrast, the orthogonal regression approach performs surprisingly badly. It seems that the geometric properties of normal Laguerre tessellations (e.g., coinciding faces, edges and vertices) favor realistic reconstructions. We think this confirms that the Laguerre tessellation is a good choice for representing polycrystalline microstructures. It is possible, of course, that the quality of the fit could be improved using non-convex cells. For example, in \cite{Alpers15}, the linear programming method was used to obtain a generalized power diagram approximation of similar data (but with far fewer grains and voxels) resulting in a fit that correctly assigned $93.8$\,\% of the voxels with non-convex cells.

\begin{figure}[htb]
\begin{center}
\subfigure[Scatter plot of cell sizes obtained from \mbox{original} and extracted cells.]{\includegraphics[width=0.45\textwidth,trim=0cm 1cm 1cm 2cm,clip]{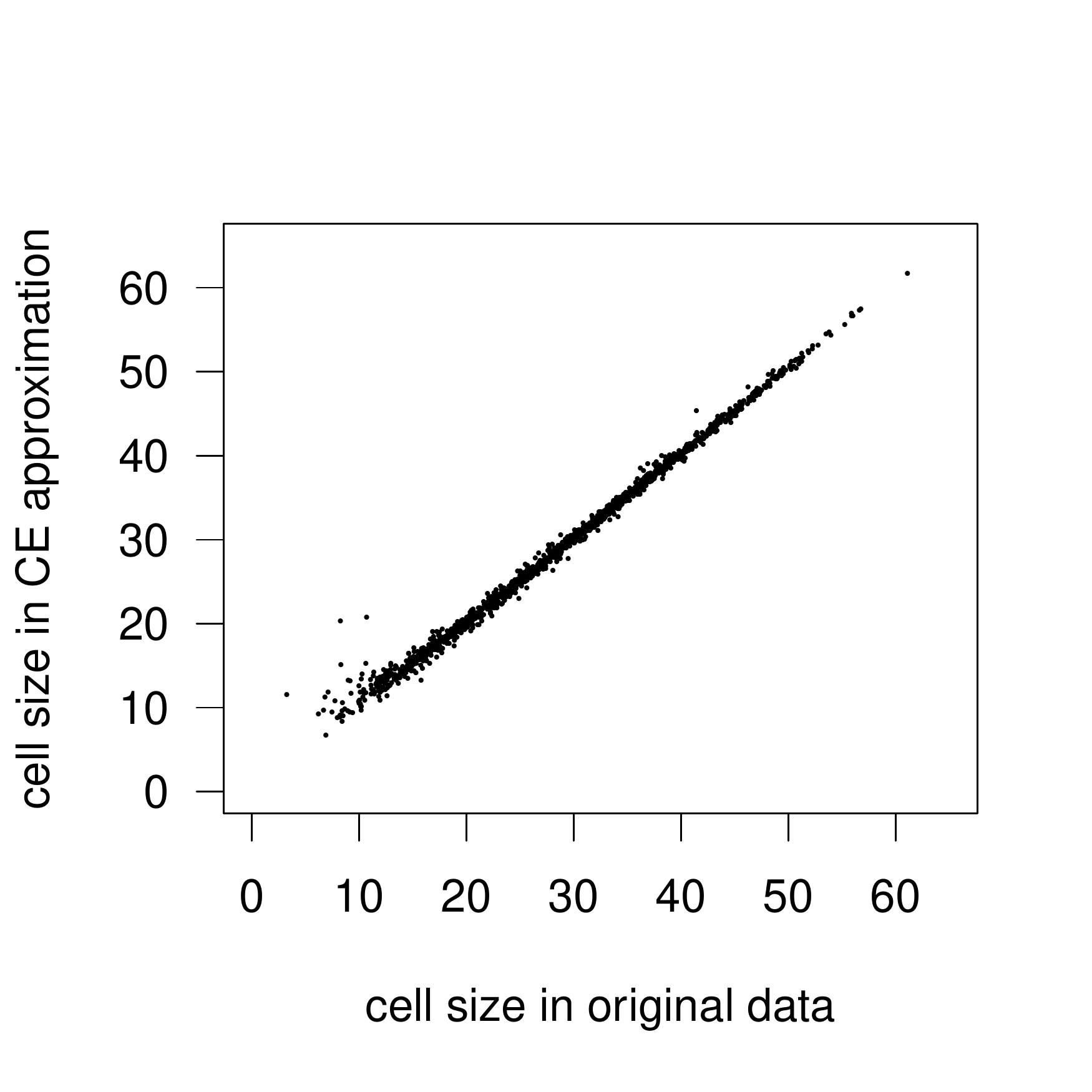}}
\subfigure[Distances between centroids of original and approximated cells, in dependence on cell sizes.]{\includegraphics[width=0.45\textwidth,trim=0cm 1cm 1cm 2cm,clip]{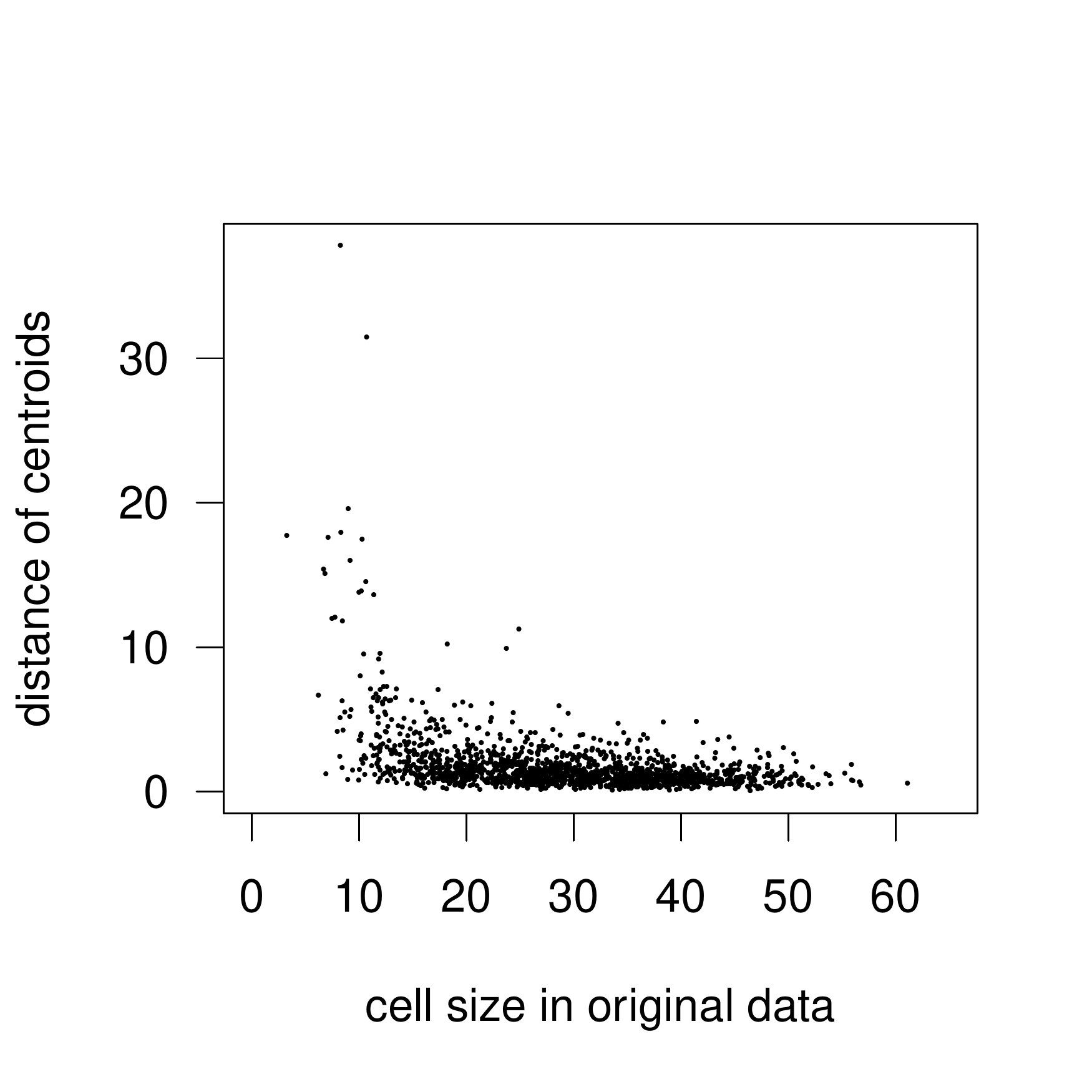}}
\caption{Experimental data: comparison of original data and cells extracted by the CE method. Cell sizes are given as the radii of their volume-equivalent spheres.}
\label{fig:experimental:cell-characteristics}
\end{center}
\end{figure}

Figure~\ref{fig:experimental:cell-characteristics}(a) shows a scatter plot of the volumes of the original grains against the volumes of the Laguerre cells (with the volumes expressed as radii of volume-equivalent spheres). The overall fit of the cell volumes is excellent. Figure~\ref{fig:experimental:cell-characteristics}(b) shows that the locations of the grains are also quite accurate. Note that the main issue with fitting seems to be small cells. This is not such a problem, however, as small cells (with equivalent radii of up to 10 voxels) are subject to image segmentation error and are, thus, not too reliable in the original data.

\section{Conclusions and outlook}\label{sec:conclusions}

In this paper, we considered the problem of approximating tomographic data by a Laguerre tessellation. We expressed this problem as an optimization problem: the generating points of the approximating tessellation need to be chosen in order to minimize the discrepancy between the tessellation and the tomographic data. We considered an interface-based discrepancy measure, instead of the volume-based discrepancies more commonly considered in the literature. This allowed us to use the CE method, a stochastic optimization method that is able to escape local minima, to fit the approximating tessellation. We then carried out numerical experiments on both artificially generated and experimentally obtained tomographic data that demonstrated the broad effectiveness of our approach.

Our method is robust and easy to implement. Thus it can be applied to fit Laguerre tessellations to almost any material with a granular or cellular structure. An obvious next step is to extend our approach to tessellations that include non-convex cells. In particular, we believe the approach and philosophy outlined in this paper can be extended to fit generalized power diagrams (see \cite{Altendorf14, Alpers15})  --- generalizations of Laguerre tessellations that can include non-convex cells. This will be the subject of a forthcoming research paper.

\section*{Funding}

This work was partially supported by the Australian Research Council under Grant DP140101956 and by the Deutsche Forschungsgemeinschaft under Grant KR 1658/4-1.

\section*{Supplemental material}

A software package including Java code and data sets can be downloaded from \url{https://github.com/stochastics-ulm-university/laguerre-approximation}.


\end{document}